\documentclass[11pt,a4paper]{article}
\usepackage[utf8]{inputenc}
\usepackage{jheppub}
\usepackage{slashed}
\usepackage{amsmath, amssymb, graphicx, physics, dsfont,array,subcaption}
\usepackage[compat=1.1.0]{tikz-feynman}  
\usepackage{orcidlink}
\usepackage{comment}
\usepackage{multirow}
\usepackage{float}
\usepackage{adjustbox}
\usepackage{placeins}

\usepackage{longtable} 
\usepackage{array}     

\usepackage[normalem]{ulem}

\allowdisplaybreaks

\usepackage{amsmath, bm}
\usepackage{graphicx}

\newcommand{\be}{\begin{equation}}
\newcommand{\ee}{\end{equation}}
\newcommand{\bes}{\begin{subequations} \begin{align} }
\newcommand{\ees}{\end{subequations}\end{align} }
\newcommand{\bea}{\begin{eqnarray}}
\newcommand{\eea}{\end{eqnarray}}

\newcommand{\eq}[1]{Eq.~\eqref{eq:#1}}

\renewcommand{\sec}[1]{Sec.~\ref{sec:#1}}

\newcommand{\fig}[1]{Fig.~\ref{fig:#1}}
\newcommand{\figs}[2]{Figs.~\ref{fig:#1} and \ref{fig:#2}}
\newcommand{\tab}[1]{Table~\ref{tab:#1}}

\def\bal#1\eal{\begin{align}#1\end{align}}

\newcounter{RSQ}

\newcounter{MSQ}

\title{
$P$--wave quarkonia in jets at the LHC}

\author[a]{Taewook Ha\orcidlink{0000-0002-5725-4007}}
\author[b]{Byungsik Hong\orcidlink{0000-0002-2259-9929}}
\author[c,*]{Daekyoung Kang\orcidlink{0000-0002-9145-6913}}
\author[b]{Soohwan Lee\orcidlink{0000-0001-9257-9643}}
\author[b,*]{Bayu Adi Nugraha Putra\orcidlink{0000-0001-9863-2839}}

\affiliation[a]{Key Laboratory of Nuclear Physics and Ion-beam Application (MOE) and Institute of Modern Physics, Fudan University, Shanghai, China 200433}
\affiliation[b]{
Department of Physics, Korea University, Seoul 02841, Korea}
\affiliation[c]{
Department of Physics, Sejong University, Seoul 05006, Korea}

\note[*]{Co-corresponding authors.}

\emailAdd{taewookha@fudan.edu.cn}
\emailAdd{bhong@korea.ac.kr}
\emailAdd{dkang@sejong.ac.kr}
\emailAdd{vince402@korea.ac.kr}
\emailAdd{bayuadinp@korea.ac.kr}

\abstract{
We present the first systematic theoretical study of the production of $P$-wave charmonium states $\chi_{cJ}$ and bottomonium states $\chi_{bJ}(nP)$ within jets at the LHC using the fragmenting jet function framework at next-to-leading order in $\alpha_s$. Our calculation resums both collinear DGLAP logarithms and threshold double logarithms in the fragmentation functions and accounts for feeddown contributions from higher quarkonium states. We observe a notable contrast between the charmonium and bottomonium sectors. For $\chi_{bJ}(nP)$, the normalized momentum-fraction distributions exhibit a nearly universal shape, with remarkably weak dependence on the choice of long-distance matrix element sets, the radial excitation ($n=1,2,3$), and the total angular momentum ($J=0,1,2$). We attribute this universality to the overwhelming dominance of the ${}^3S_1^{[8]}$ fragmentation channel, while feeddown contributions from higher bottomonium states remain small. In contrast, the shape of $\chi_{cJ}$ distributions exhibits a stronger dependence on the matrix-element set, reflecting the different relative contributions of the ${}^3S_1^{[8]}$ and ${}^3P_J^{[1]}$ channels.
We provide predictions for the $\chi_{cJ}$ and $\chi_{bJ}(nP)$ distributions for the CMS and LHCb kinematic selections across a range of jet transverse momenta. We also compare our predictions with \textsc{Pythia}~8 simulations for selected $P$-wave quarkonium states.}

\begin{document}
\maketitle

\section{Introduction}
\label{sec:1}

Ever since the discovery of the $J/\psi$, heavy quarkonium has provided an invaluable laboratory for exploring strong interactions across both perturbative and nonperturbative regimes.
Although nonrelativistic QCD (NRQCD)~\cite{Bodwin:1994jh} has proven broadly successful in describing quarkonium production, the underlying production mechanisms are not yet fully understood.
Because inclusive cross sections lack sufficient sensitivity to distinguish individual production channels, available extractions of the corresponding long-distance matrix elements (LDMEs) show large uncertainties and significant tensions with one another~\cite{Butenschoen:2011yh,Gong:2012ug,Han:2014jya,Zhang:2014ybe,Shao:2014yta,Bodwin:2015iua,Feng:2018ukp,Brambilla:2021abf,Butenschoen:2022qka,Cho:1995vh,Cho:1995ce,Ma:2010vd,Baranov:2015yea,Baranov:2023ckv,Zhang:2014coi}.
This ambiguity underscores the need for more differential observables. In particular, studying momentum-fraction distributions of quarkonium within jets has emerged as an effective strategy to disentangle the contributions from different production channels~\cite{Baumgart:2014upa,Bain:2016rrv,Kang:2017yde}.

On the experimental side, measurements of quarkonium production within jets have so far focused primarily on $S$-wave quarkonium.
The first such study was reported by the LHCb Collaboration for $J/\psi$ mesons in $pp$ collisions~\cite{LHCb:2017llq}. The CMS Collaboration measured jet fragmentation distributions for prompt and nonprompt $J/\psi$ in $pp$ collisions~\cite{CMS:2019ebt}, as well as in heavy-ion collisions~\cite{CMS:2021puf} to investigate medium-induced modifications of quarkonium production within jets. LHCb later measured the radial excitation $\psi(2S)$ and the exotic state $X(3872)$ within fully reconstructed jets~\cite{LHCb:2024ybz}, covering a range of jet $p_T$ and quarkonium $p_T$.
The ALICE Collaboration also measured the transverse-momentum fraction carried by $J/\psi$ mesons in charged-particle jets~\cite{ALICE:2026ywy}. More recently, the CMS Collaboration reported the first measurement of $\Upsilon(nS)$ production within jets~\cite{CMS:2026hnn}.

In general-purpose Monte Carlo event generators, NRQCD fragmentation-function (FF) based descriptions of quarkonium production have recently been incorporated into parton showers.
Specifically, a timelike parton shower incorporating splitting kernels derived from FFs was implemented in \textsc{Pythia}~8~\cite{Cooke:2023ldt}, enabling full event-level simulations of quarkonium within jets.
Similarly, a fully automated quarkonium shower framework was developed for the angularly ordered shower in \textsc{Herwig}~7~\cite{Masouminia:2025kec}.
Applications of quarkonium showers to this observable have also been presented~\cite{LHCb:2024ybz,ValenciaPalomo:2025ljc,CMS:2026hnn}.

Theoretically, quarkonium production inside jets can be factorized using effective field theories and systematically formulated within the fragmenting jet function (FJF) framework~\cite{Procura:2009vm,Jain:2011xz,Baumgart:2014upa,Kang:2016ehg}. This formalism enables predictions of quarkonium polarization~\cite{Kang:2017yde} and transverse-momentum dependence~\cite{Kang:2017glf,Copeland:2023wbu,Copeland:2025vop}, as well as the momentum fraction carried by the quarkonium within the jet, which has been widely measured experimentally. Within the FJF approach, theoretical challenges arise in the threshold region where the quarkonium carries most of the jet momentum. In this regime, the restricted phase space for real emissions generates large threshold logarithms in the short-distance coefficients (SDCs), requiring threshold resummation for reliable theoretical predictions~\cite{Chung:2025gjk,Chung:2026mii}.

Phenomenological studies of the momentum fraction distribution at the LHC based on the FJF framework were first performed for $J/\psi$ in Ref.~\cite{Bain:2016clc} and later for $\psi(2S)$ in Refs.~\cite{Copeland:2025osx}, both at leading-logarithmic (LL) accuracy. These studies were subsequently extended to next-to-leading order (NLO) plus LL accuracy with threshold resummation for both states in Ref.~\cite{Wang:2025drz}. 
These studies demonstrated notable sensitivity to the choice of LDME sets, highlighting the potential of jet substructure observables to probe the underlying quarkonium production dynamics. While gluon-initiated jets dominate at the LHC, quark-initiated jets~\cite{AbdulKhalek:2021gbh,Accardi:2012qut,Ee:2025scz,Zhu:2021xjn,Chu:2022jgs,Kang:2014qba,Kang:2013nha} play a prominent role at the Electron-Ion Collider (EIC). 
Consequently, quarkonium-in-jet measurements at the EIC offer a complementary probe of charmonium production dynamics~\cite{Wang:2026dul,EICph:2026hgp}.

In contrast to charmonium, a recent study of $S$-wave bottomonium production, $\Upsilon(nS)$, found that their shape of the momentum fraction distribution is remarkably insensitive to the choice of LDME sets~\cite{Ha:2026mct}. This behavior was attributed to the overwhelming dominance of the ${}^3S_1^{[8]}$ fragmentation channel, leading to a universal normalized distribution across the $n=1,2,3$ radial excitation states.

While previous studies have focused on $S$-wave states such as $J/\psi$, $\psi(2S)$, and $\Upsilon(nS)$~\cite{Wang:2025drz,Wang:2026dul,Ha:2026mct}, a comprehensive understanding of their production remains challenging because of significant feeddown contributions. In particular, feeddown transitions from the $\chi_{cJ}$ and $\chi_{bJ}(nP)$ states constitute a sizable fraction of the observed prompt $J/\psi$ and inclusive $\Upsilon(nS)$ cross sections. A systematic investigation of $\chi_{cJ}$ and $\chi_{bJ}(nP)$ production inside jets is therefore important for disentangling these contributions and improving our understanding of the underlying production dynamics.

In addition to providing direct access to feeddown sources, studying $P$-wave quarkonium production offers a simpler channel structure than the $S$-wave sector. In NRQCD, the production of $S$-wave states involves four leading contributions in the $v$ expansion: the color singlet (CS) ${}^3S_1^{[1]}$ channel and three color octet (CO) channels, ${}^3S_1^{[8]}$, ${}^1S_0^{[8]}$, and ${}^3P_J^{[8]}$. The interplay among these channels makes it difficult to disentangle their individual contributions and determine the corresponding LDMEs precisely. In contrast, the production of $P$-wave quarkonium states is governed at LO in $v$ by only two channels: the ${}^3P_J^{[1]}$ and the ${}^3S_1^{[8]}$. This simpler channel structure facilitates a more direct examination of the relative contributions of the CS and CO channels, providing a useful laboratory for studying the quarkonium production mechanism.

In this paper, we present a comprehensive phenomenological study of the production of $P$-wave charmonium states $\chi_{cJ}$ and bottomonium states $\chi_{bJ}(nP)$ within jets at the LHC. We consider \emph{prompt} $\chi_{cJ}$ production, and \emph{inclusive} $\chi_{bJ}(nP)$ production \footnote{Here, \emph{prompt} production includes direct production and feeddown from higher charmonium states, while excluding contributions from $B$-meson decays. For bottomonium, we use \emph{inclusive} to denote direct production together with feeddown from higher bottomonium states; contributions from $B$-meson decays are not relevant}.
We find a clear contrast between the two sectors: while the normalized momentum-fraction distributions for $\chi_{cJ}$ exhibit stronger dependence on the LDME set, those for $\chi_{bJ}(nP)$ show nearly universal behavior across LDME sets and radial excitations, owing to the overwhelming dominance of the ${}^3S_1^{[8]}$ channel.

The structure of this paper is as follows. \sec{2} outlines our setups, including the factorization formalism, the incorporation of threshold-resummed FFs, the NRQCD matrix elements used in our analysis, and the setup for the \textsc{Pythia}~8 simulations. In \sec{3}, we present our numerical results for $P$-wave quarkonium production within jets, including the dependence on quark flavor, LDME sets, and jet transverse momentum. We also discuss the impact of the CMS and LHCb kinematic acceptances and compare our predictions with \textsc{Pythia}~8 simulations. We summarize our findings and conclude in \sec{4}.
\section{Setup and input parameters} \label{sec:2}

\subsection{Theoretical framework} \label{sec:theory}

We employ the FJF framework~\cite{Kang:2016ehg}, following our previous study of $S$-wave quarkonium production within jets. We therefore summarize only the ingredients relevant for the present $P$-wave analysis. 

The production of heavy quarkonium within a high-$p_T$ jet involves three well-separated dynamical scales:
\begin{equation}
    \mu_H \sim p_T \gg \mu_J \sim p_T R \gg \mu_0 \sim 2m_Q,
    \label{eq:scales}
\end{equation}
where $\mu_H$, $\mu_J$, and $\mu_0$ are the hard, jet, and quarkonium mass scales, respectively. The kinematics are parameterized by two momentum fractions:
\begin{equation}
    z \equiv \frac{p_T}{p_T^i}, \qquad z_H \equiv \frac{p_T^H}{p_T},
    \label{eq:z_definitions}
\end{equation}
where $p_T^i$ is the transverse momentum of the initiating parton $i$, $p_T$ is the jet transverse momentum, and $p_T^H$ is that of the observed quarkonium state $H = \chi_{QJ}$ ($Q=c,b$, $J=0,1,2$).

When the jet scale satisfies $p_T R \gg 2m_Q$, the differential yield can be systematically factorized within Soft-Collinear Effective Theory (SCET)~\cite{Bauer:2000yr,Bauer:2001ct} as
\begin{equation}
\frac{d\sigma^H}{dp_T\,d\eta\,dz_H}
=
\sum_i
\frac{d\sigma^i}{dp_T^i\,d\eta}
\otimes\,
\mathcal{G}_i^H(z,z_H,p_TR,\mu),
\label{eq:factorization}
\end{equation}
where $d\sigma^i$ denotes the inclusive production cross section for the initiating parton $i$, differential in $p_T^i$ and the jet pseudorapidity $\eta$, and includes the convolution with the incoming proton PDFs. The semi-inclusive FJF, $\mathcal{G}_i^H$, matches onto single-hadron FFs via
\begin{equation}
    \mathcal{G}_i^H(z, z_H, p_T R, \mu) = \sum_j \mathcal{J}_{ij}\left(z, \frac{z_H}{\xi}, p_T R, \mu\right) \otimes D_{j\to H}(\xi, \mu) + \mathcal{O}\left(\frac{m_Q^2}{p_T^2 R^2}\right),
    \label{eq:matching_FJF_FF}
\end{equation}
where the perturbative coefficients $\mathcal{J}_{ij}$ are evaluated through NLO in $\alpha_s$ and account for radiation inside and outside the jet of radius $R$~\cite{Kang:2016ehg}.

NRQCD factorization~\cite{Bodwin:1994jh} allows the FF to be expressed as a sum of products of SDCs and LDMEs:
\begin{equation}
    D_{j \to \chi_{QJ}}(z_H, \mu_0) = \sum_n d_{j \to Q\bar{Q}[n]}(z_H, \mu_0) \, \langle \mathcal{O}^{\chi_{QJ}}(n) \rangle,
    \label{eq:NRQCDmatching}
\end{equation}
where $d_{j\to Q\bar Q[n]}$ are perturbatively calculable SDCs, while the LDMEs $\langle {\cal O}^{\chi_{QJ}}(n)\rangle$ encode the corresponding long-distance transition into the physical quarkonium state. For a $P$-wave quarkonium state $\chi_{QJ}$, with $Q=c,b$, the leading contributions in the NRQCD velocity expansion arise from the ${}^3S_1^{[8]}$ and ${}^3P_J^{[1]}$ channels. Additional Fock-state contributions enter at higher orders in the $v$ expansion and are not included in the present analysis. The SDCs for both channels are computed through NLO in $\alpha_s$~\cite{Ma:2013yla}.

Large logarithms generated by the separation of physical scales in Eq.~\eqref{eq:scales} are handled through a two-stage renormalization-group evolution at LL accuracy. First, timelike DGLAP equations evolve the FFs from $\mu_0$ to the jet scale $\mu_J$, summing $\ln(\mu_J/\mu_0)$ terms. Second, evolution of the FJF from $\mu_J$ to the $\mu_H$ resums logarithms associated with the jet radius $R$.

In quarkonium fragmentation near the threshold limit $z_H \to 1$, the restricted phase-space constraints on real radiation produce large threshold double logarithms of the form $\alpha_s^k \, \frac{\ln^{2k-1}(1-z_H)}{1-z_H}$. These logarithms lead to singular behavior of the fixed-order SDCs for ${}^3S_1^{[8]}$ and ${}^3P_J^{[1]}$ near the threshold. To improve the perturbative behavior in this region, we implement threshold resummation at LL accuracy~\cite{Chung:2024jfk,Chung:2026mii}.

\subsection{LDME sets}\label{sec:LDME}

Heavy-quark spin symmetry relates the LDMEs within a given $\chi_{QJ}$ multiplet. For the relevant channels, the relations can be written as

\begin{equation}
\begin{aligned}
\langle \mathcal{O}^{\chi_{QJ}}({}^3S_1^{[8]})\rangle
&=
(2J+1)\,
\langle \mathcal{O}^{\chi_{Q0}}({}^3S_1^{[8]})\rangle
\left[1+\mathcal{O}(v_Q^2)\right],
\\
\langle \mathcal{O}^{\chi_{QJ}}({}^3P_J^{[1]})\rangle
&=
(2J+1)\,
\langle \mathcal{O}^{\chi_{Q0}}({}^3P_0^{[1]})\rangle
\left[1+\mathcal{O}(v_Q^2)\right].
\end{aligned}
\label{eq:HQSS}
\end{equation}

Thus, it is sufficient to specify the two independent LDMEs for $\chi_{Q0}$ for each heavy-quark flavor.

\begin{table}[thb]
	\centering
	\begin{tabular}{cccc}
		\hline
        \multirow{2}{*}{\textbf{State}} 
		& \multirow{2}{*}{\phantom{Empty}} 
		& $\langle \mathcal{O}(^{3}S^{[8]}_{1}) \rangle$ 
		& $\langle \mathcal{O}(^{3}P^{[1]}_{0}) \rangle / m_c^2$ \\ 
		&
		& $\times10^{-2}~\text{GeV}^3$ 
		& $\times10^{-2}~\text{GeV}^3$ \\ \hline\hline
		\multirow{2}{*}{$\chi_{c0}$}
		& Brambilla \textit{et al.}~\cite{Brambilla:2021abf}
		& $0.157 \pm 0.007$ & $3.63$ \\ \cline{2-4}
		
		& Bodwin \textit{et al.}~\cite{Bodwin:2015iua}
		& $0.574 \pm 0.131$ & $3.53 \pm 1.08$ \\ \hline\hline
	\end{tabular}
    \caption{Representative $\chi_{c0}$ LDME sets used in the calculation, provided in units of $10^{-2}\,\text{GeV}^3$ and $m_c=1.5$~GeV.}
	\label{tab:chic0_ME}
\end{table}

\begin{table}[H]
	\centering
	\begin{tabular}{cccc}
		\hline
		\multirow{2}{*}{\textbf{State}} 
		& \multirow{2}{*}{\phantom{Empty}} 
		& $\langle \mathcal{O}(^{3}S^{[8]}_{1}) \rangle$ 
		& $\langle \mathcal{O}(^{3}P^{[1]}_{0}) \rangle / m_b^2$ \\ 
		& & $\times10^{-2}~\text{GeV}^3$ & $\times10^{-2}~\text{GeV}^3$ \\ \hline\hline
		
		\multirow{2}{*}{$\chi_{b0}(1P)$}
		& Han \textit{et al.}~\cite{Han:2014kxa}
		& $0.73 \pm 0.09$ & $1.73$ 
        \\ \cline{2-4}
		& Brambilla \textit{et al.}~\cite{Brambilla:2021abf}
		& $1.44 \pm 0.20$ & $1.55$ 
        \\ \hline
		
		\multirow{2}{*}{$\chi_{b0}(2P)$}
		& \multirow{2}{*}{Same as $1P$}
		& $1.08 \pm 0.14$ & $1.73$ 
        \\ \cline{3-4}
        & 
		& $1.71 \pm 0.24$ & $1.86$ 
        \\ \hline
		
		\multirow{2}{*}{$\chi_{b0}(3P)$}
        & \multirow{2}{*}{Same as $1P$}
		& $1.08 \pm 0.14$ & $1.73$ 
        \\ \cline{3-4}
		& 
		& $1.71 \pm 0.24$ & $1.86$ 
        \\ \hline\hline
	\end{tabular}
    \caption{Representative $\chi_{b0}(nP)$ ($n=1,2,3$) LDME sets used in the calculation, provided in units of $10^{-2}\,\text{GeV}^3$ and $m_b=4.75$~GeV.}
    \label{tab:chib0_ME}
\end{table}

Among the various sets available in the literature~\cite{Cho:1995vh,Cho:1995ce,Brambilla:2021abf,Bodwin:2015iua,Shao:2014yta,Gong:2012ug,Feng:2018ukp,Ma:2010vd,Baranov:2015yea,Baranov:2023ckv,Zhang:2014coi}, we adopt two representative LDME sets for each of the charm and bottom sectors, as listed in \tab{chic0_ME} and \tab{chib0_ME}. 
In the tables, only the matrix elements for the $J=0$ states are listed. 
The corresponding LDMEs for $J=1$ and $2$ are obtained using heavy-quark spin symmetry relations in \eq{HQSS}.
These benchmark sets are chosen to span a representative range of values for the two relevant LDMEs, with the values from other parameterizations generally lying within the range spanned by these two choices.\footnote{In Brambilla \textit{et al.}~\cite{Brambilla:2021abf}, there are three different values of $\langle \mathcal{O}({}^3S_1^{[8]})\rangle$ for $\chi_{c0}$ depending on the fitting setup, and the smallest value is used to maximize the contrast with the other set shown in \tab{chic0_ME}.}
The LDMEs used for the feeddown contributions from higher quarkonium states are summarized separately in Appendix~\ref{app:feeddownLDME}.

Several notable features of the representative LDME sets in \tab{chic0_ME} and \tab{chib0_ME} are worth highlighting. 
First, all LDMEs in those tables are positive, whereas some CO LDMEs for $S$-wave quarkonia can be either positive or negative depending on the LDME set.
Second, the rescaled CS LDME, $\langle \mathcal{O}({}^3P_0^{[1]})\rangle/m_Q^2$, is larger than the CO LDME, $\langle \mathcal{O}({}^3S_1^{[8]})\rangle$, in both the charmonium and bottomonium sectors. This hierarchy is particularly pronounced for charmonium, where the CS LDME exceeds the CO LDME by factors of approximately $6$--$23$, whereas the two are of comparable magnitude for bottomonium.
Finally, when comparing the two benchmark sets within each sector, the CS LDMEs remain relatively stable, with variations of only a few percent for $\chi_{c0}$ and around $10\%$ for $\chi_{b0}$. In contrast, the CO LDMEs show substantially larger variations. In particular, $\langle \mathcal{O}({}^3S_1^{[8]})\rangle$ differs by a factor of approximately $3.7$ between the Brambilla and Bodwin sets for $\chi_{c0}$, and by a factor of approximately $1.6$--$2.0$ between the Han and Brambilla sets across the $nP$ bottomonium states.
These differences between LDME sets are reflected in the normalized $z_H$ distributions discussed in Sec.~\ref{sec:3}.

\subsection{Simulation setup}\label{sec:PYTHIA}

We perform Monte Carlo simulations of proton--proton collisions at $\sqrt{s}=13$~TeV to study $P$-wave quarkonium production within jets, and use selected states for comparison with our theoretical predictions. Events are generated using \textsc{Pythia}~8 version~8.316~\cite{pythia:2006,pythia:2008,pythia:2015,pythia:2022} with the Monash 2013 tune~\cite{Skands:2014pea} and analyzed using the \textsc{Rivet} framework~\cite{rivet:2024}. Quarkonium production within the parton shower is handled by \textsc{OniaShower}, which incorporates NRQCD-based FFs~\cite{Cooke:2023ldt}. We use the default LDME values provided by \textsc{OniaShower}. Jets are reconstructed using the anti-$k_T$ algorithm with a distance parameter of $R=0.4$ for the CMS selection and $R=0.5$ for the LHCb selection. The muon acceptance requirements described in Appendix~\ref{app:mucut} are applied separately for the CMS and LHCb configurations.

\section{Numerical results}
\label{sec:3}

\begin{figure}[tb]
    \centering
    \includegraphics[width=0.7\textwidth]{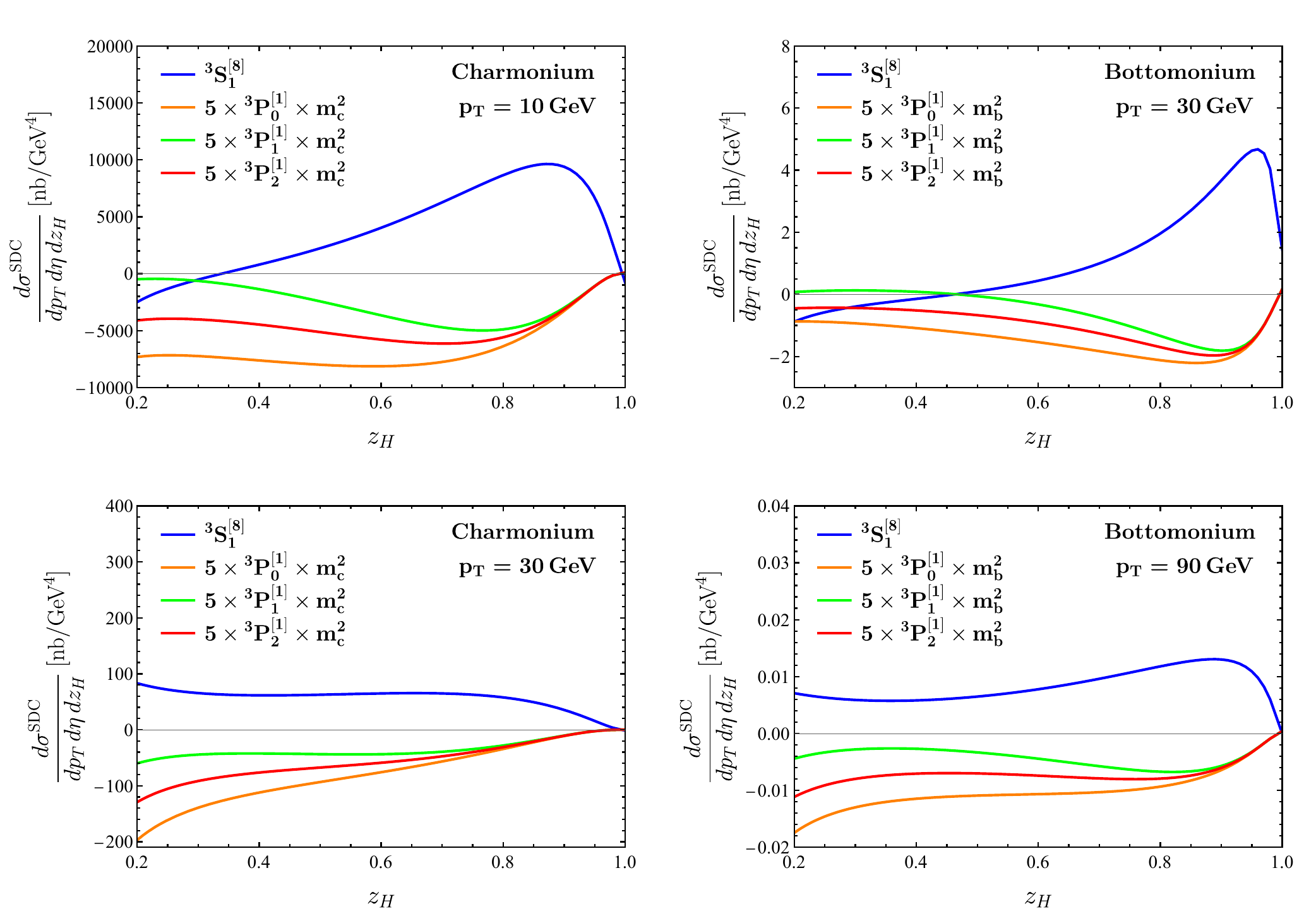}
    \caption{
    SDCs for ${Q\bar Q[n]}$ pair production for $n={}^3S_1^{[8]}$ and ${}^3P_J^{[1]}$ channels as a function of $z_H$ at different $p_T$, with the jet pseudorapidity $\eta=0$ for charmonium (left) and bottomonium (right). 
    The ${}^3P_J^{[1]}$ SDCs are multiplied by a factor of 5 for visibility and scaled by $m_Q^2$ to match the mass dimension, with $m_c = 1.5\text{ GeV}$ and $m_b = 4.75\text{ GeV}$.
    }
    \label{fig:sdcs}
\end{figure}

In this section, we present our numerical results for the $z_H$ distributions of $P$-wave charmonium and bottomonium. First, we show the production cross sections of the $Q\bar{Q}[n]$ pairs, the SDCs, in the individual channels ($n = {}^3S_1^{[8]}, {}^3P_J^{[1]}$) and discuss their features. We then discuss the full distributions for charmonium and bottomonium separately in the following subsections.

\fig{sdcs} shows the individual SDCs as functions of $z_H$ at representative $p_T$, for charmonium (left) and bottomonium (right). 
The ${}^3P_J^{[1]}$ SDCs are generally smaller in magnitude than those of the ${}^3S_1^{[8]}$ channel over the kinematic region considered.
This is consistent with the fact that the ${}^3S_1^{[8]}$ channel contributes already at LO in $\alpha_s$ via gluon fragmentation, whereas the ${}^3P_J^{[1]}$ channel only begins at NLO.
The ${}^3S_1^{[8]}$ channel exhibits a pronounced enhancement at large $z_H$ for relatively low $p_T$. As the $p_T$ increases, this peak structure gradually becomes less pronounced and broader, with a substantial fraction of the distribution extending away from the large-$z_H$ region.

In contrast, the ${}^3P_J^{[1]}$ SDCs are negative over most of the kinematic region and exhibit broader shapes than those of the ${}^3S_1^{[8]}$ channel.
Their $J$ dependence does not exhibit a simple hierarchy, with the relative size and shape of the three ${}^3P_J^{[1]}$ contributions depending on $z_H$ and the transverse-momentum scale. This nontrivial $J$ dependence of the SDCs implies that the relative production rates of $\chi_{Q0}$, $\chi_{Q1}$, and $\chi_{Q2}$ need not be proportional to $2J+1$, even when this channel is subleading.

\subsection{Charmonium}

We now examine the resulting $z_H$ distributions for \emph{prompt} $\chi_{cJ}$ production, starting with direct production and then including feeddown from higher charmonium states.

Regarding the direct contribution, the Brambilla LDME set features a relatively small ${}^3S_1^{[8]}$ LDME, allowing the ${}^3P_J^{[1]}$ channel to appreciably alter the direct $z_H$ distribution away from the characteristic ${}^3S_1^{[8]}$ shape. In contrast, the Bodwin set exhibits a larger ${}^3S_1^{[8]}$ LDME and leads the direct distribution to more closely follow the ${}^3S_1^{[8]}$ profile.

Turning to the feeddown contribution, the $\psi(2S)\to\chi_{cJ}$ channel in the Brambilla LDME set is most prominent for $J=0$ and decreases in significance with increasing $J$. This behavior arises because the direct production rate grows with $J$, primarily due to the factor $2J+1$ in the spin-symmetry relations, whereas the branching fractions $\mathcal{B}(\psi(2S)\to\chi_{cJ})$ remain comparable across $J=0,1,2$.
In contrast, for the Bodwin LDME set, the suppressed $\psi(2S)$ production rate relative to direct $\chi_{cJ}$ yields makes the feeddown contribution negligible for all three $J$ states.

In view of future comparisons with experimental data, we consider proton-proton collisions at a center-of-mass energy of $\sqrt{s}=13~\text{TeV}$ and take into account the kinematic ranges and acceptances of the CMS and LHCb experiments.
For CMS, we consider the jet pseudorapidity range $|\eta|<1.5$ and use the jet radius parameter $R=0.4$, while for LHCb we consider $2.5<\eta<4.0$ and $R=0.5$.
$P$-wave quarkonia are reconstructed through the cascade decay $\chi_{QJ}\to H_S+\gamma$, followed by $H_S\to\mu^+\mu^-$, where $H_S=J/\psi$ for charmonium and $H_S=\Upsilon$ for bottomonium. 
We adopt their muon-acceptance requirements, following
CMS~\cite{CMS:2021znk,CMS:2021mzx} and LHCb~\cite{LHCb:2017llq}. The impact of the muon-acceptance requirements on the $z_H$ distributions is discussed in Appendix~\ref{app:mucut}.

\begin{figure}[thb]
    \centering
    \includegraphics[width=0.7\textwidth]{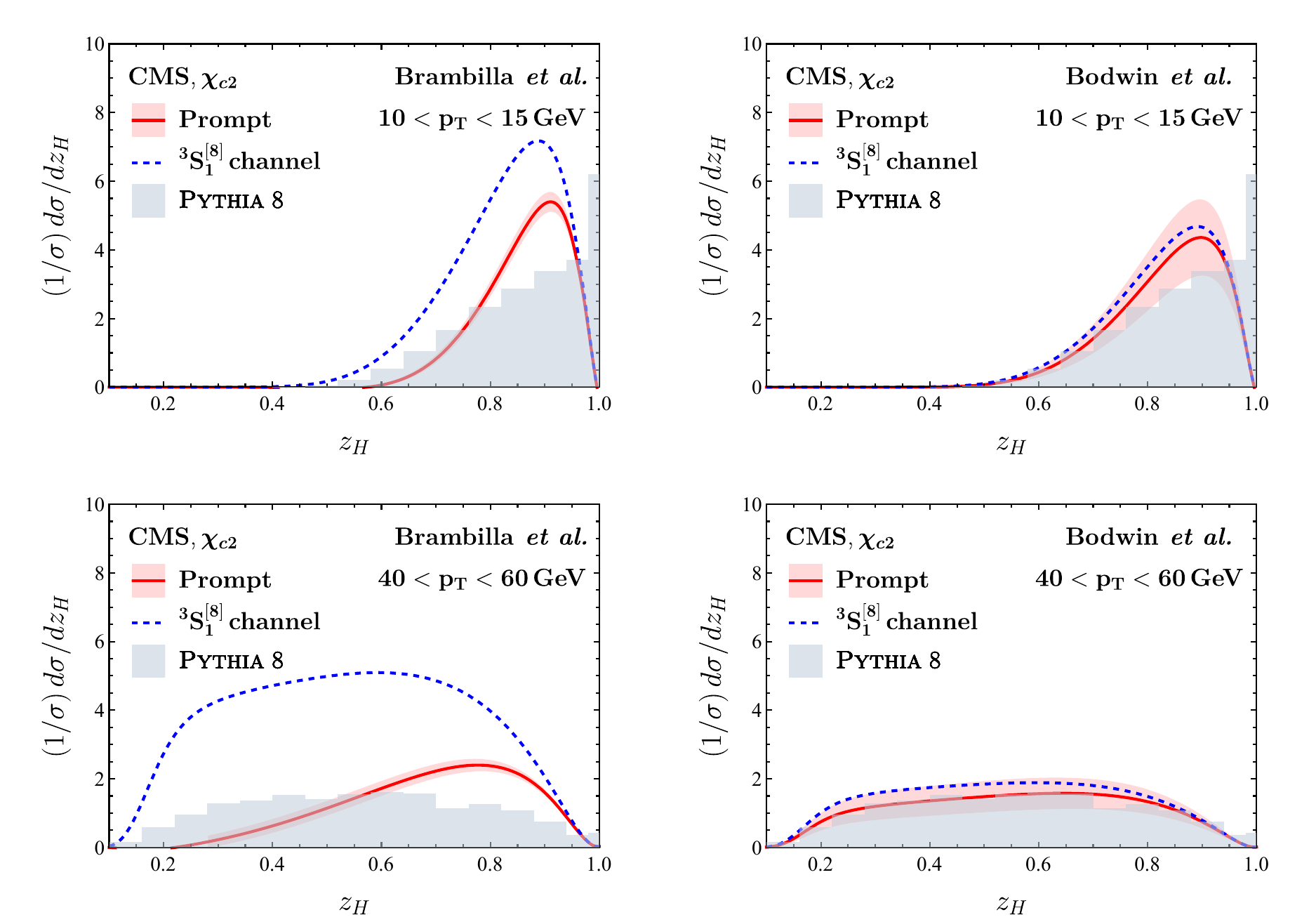}
    \caption{
    Normalized $z_H$ distributions for prompt $\chi_{c2}$ production with CMS kinematic selection.
    The upper and lower panels correspond to $10 < p_T < 15~\text{GeV}$ and $40 < p_T < 60~\text{GeV}$, respectively, while the left and right panels show the results for the two LDME sets.
    The \textsc{Pythia}~8 results are included for comparison.
    }
    \label{fig:chic2_CMS_rep}
\end{figure}

\begin{figure}[h]
    \centering
    \includegraphics[width=0.7\textwidth]{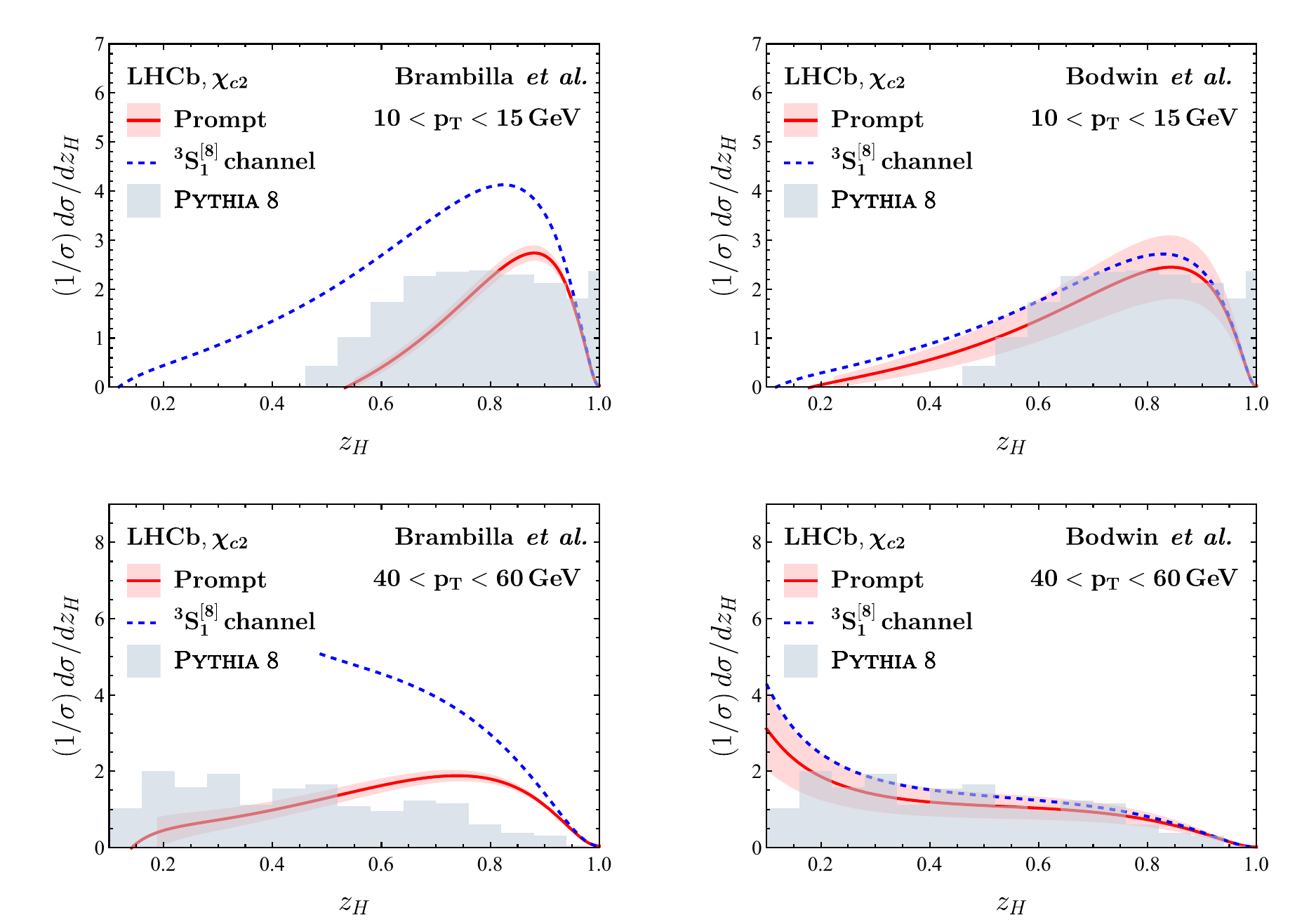}
    \caption{
    Normalized $z_H$ distributions for prompt $\chi_{c2}$ production under LHCb kinematic selections, using the same $p_T$ bins and LDME sets as in \fig{chic2_CMS_rep}.
    }
    \label{fig:chic2_LHCb_rep}
\end{figure}

Next, we examine the $z_H$ distributions for prompt $\chi_{c2}$ production, taking $J=2$ as a representative case. Results are presented for two $p_T$ bins, $10 < p_T < 15\text{ GeV}$ and $40 < p_T < 60\text{ GeV}$, using the two LDME sets in \tab{chic0_ME}.
\figs{chic2_CMS_rep}{chic2_LHCb_rep} display the results for the CMS and LHCb kinematic selections, respectively.
The results for the other $p_T$ bins and $J$ states are presented in the Appendices~\ref{app:zhCMS} and \ref{app:zhLHCb}.
The uncertainty bands shown in the figures include the propagated uncertainties of the LDMEs.
The distributions obtained with the CMS and LHCb selections exhibit similar overall shapes, but the acceptance effects differ between the two experiments. The acceptance suppression is more pronounced at lower $p_T$. In particular, the CMS selection imposes stronger suppression in the low-$z_H$ region, whereas the LHCb selection retains a larger fraction of the distribution.

We also compare our predictions with the \textsc{Pythia}~8 results for the CMS and LHCb selections. For the CMS selection, the \textsc{Pythia}~8 result better agree with the theoretical prediction using the Bodwin LDME set within its uncertainties over most of the $z_H$ range in the low-$p_T$ bin ($10 < p_T < 15~\text{GeV}$). Near the threshold, Pythia gives finite results with a sharp rise. This difference may be related to the different treatments of threshold effects, which are resummed in our calculation, forcing our distribution to vanish at the threshold but not treated in the same way in \textsc{Pythia}~8. In the high-$p_T$ bin ($40 < p_T < 60~\text{GeV}$), both our theoretical predictions and the \textsc{Pythia}~8 result are distributed toward lower $z_H$ values, with particularly good agreement for the prediction using the Bodwin LDME set over the entire $z_H$ range. For the LHCb selection, the low-$p_T$ results exhibit a similar difference near the threshold, while the high-$p_T$ bin result agrees well with the prediction using the Bodwin LDME set across the entire $z_H$ range except for the lowest-$z_H$ bin.

Turning to the theoretical predictions, the $z_H$ distribution exhibits a pronounced peak at large $z_H$ in the low-$p_T$ bin, while it broadens and shifts toward smaller $z_H$ in the high-$p_T$ bin. This behavior is observed for both LDME sets.
In both experiments, the relative contribution of the ${}^3S_1^{[8]}$ channel is more pronounced at lower $p_T$, suggesting that this kinematic region offers greater sensitivity to the corresponding LDME.
For the Bodwin set, the ${}^3S_1^{[8]}$ contribution accounts for nearly the entire distribution.

The normalized distributions exhibit qualitatively similar overall shapes for the two LDME sets, but some differences remain. In particular, the Brambilla set produces a somewhat sharper peak at large $z_H$ and a more rapid decrease toward smaller $z_H$. These differences indicate that the normalized $z_H$ distribution retains some sensitivity to the choice of LDME set even after feeddown contributions are included.
The two LDME sets can be clearly discriminated if the absolute normalization of the distribution is known. The Bodwin set yields cross sections up to four times larger than those of the Brambilla set, primarily due to the difference in their ${}^3S_1^{[8]}$ LDME values.

\begin{figure}[t]
    \centering
    \includegraphics[width=0.7\textwidth]{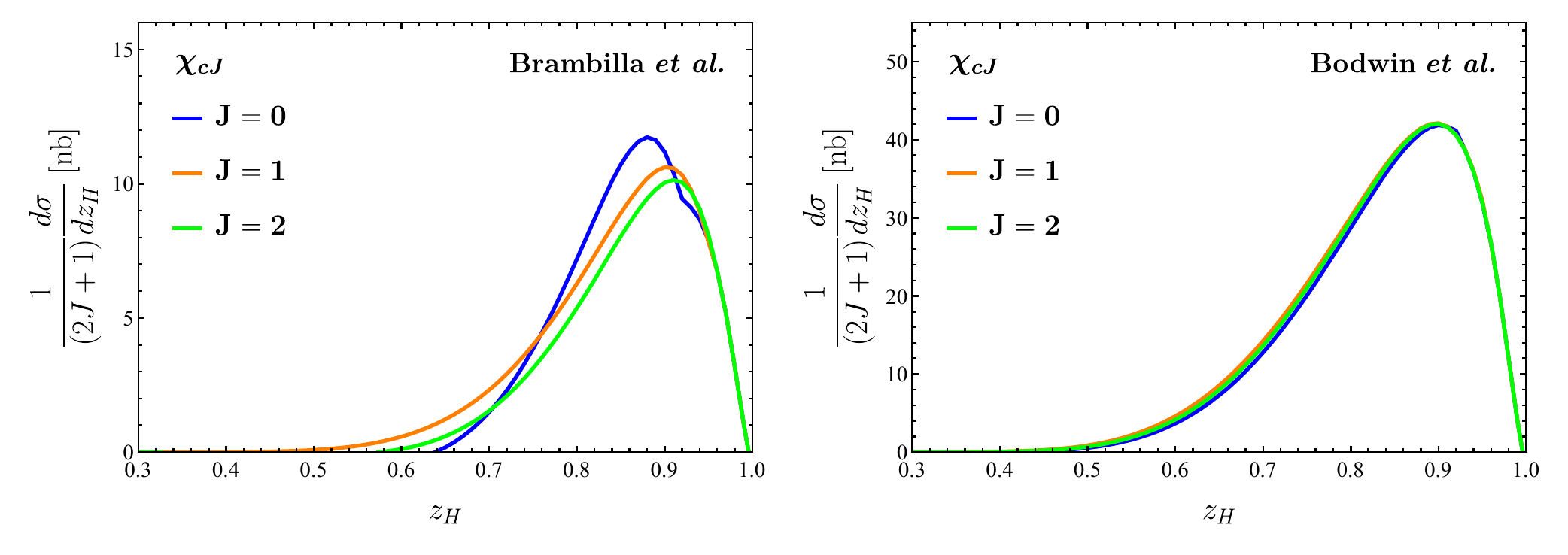}
    \caption{Comparison of the $z_H$ distributions for $\chi_{cJ}$ with $J=0,1,2$, divided by the factor $2J+1$, under the CMS kinematic selection $10 < p_T < 15~\text{GeV}$ and $|\eta|<1.5$.}
    \label{fig:J_dependence}
\end{figure}

We further investigate the $J$ dependence using the $J$-weighted distribution, $\tfrac{1}{2J+1}\tfrac{d\sigma}{dz_H}$, as shown in \fig{J_dependence}. Although the $\chi_{cJ}$ LDMEs satisfy the heavy-quark spin-symmetry relation in \eq{HQSS} at LO in the velocity expansion, the weighted distributions need not be identical because the SDCs for the ${}^3P_J^{[1]}$ channel have nontrivial $J$ dependence and break the $J$ degeneracy. For the Bodwin set, the three $\chi_{cJ}$ distributions in the figure are nearly degenerate, whereas a notable $J$ dependence is observed for the Brambilla set.

\subsection{Bottomonium}

\begin{figure}[t]
    \centering
    \begin{adjustbox}{center}
    \includegraphics[width=0.7\textwidth]{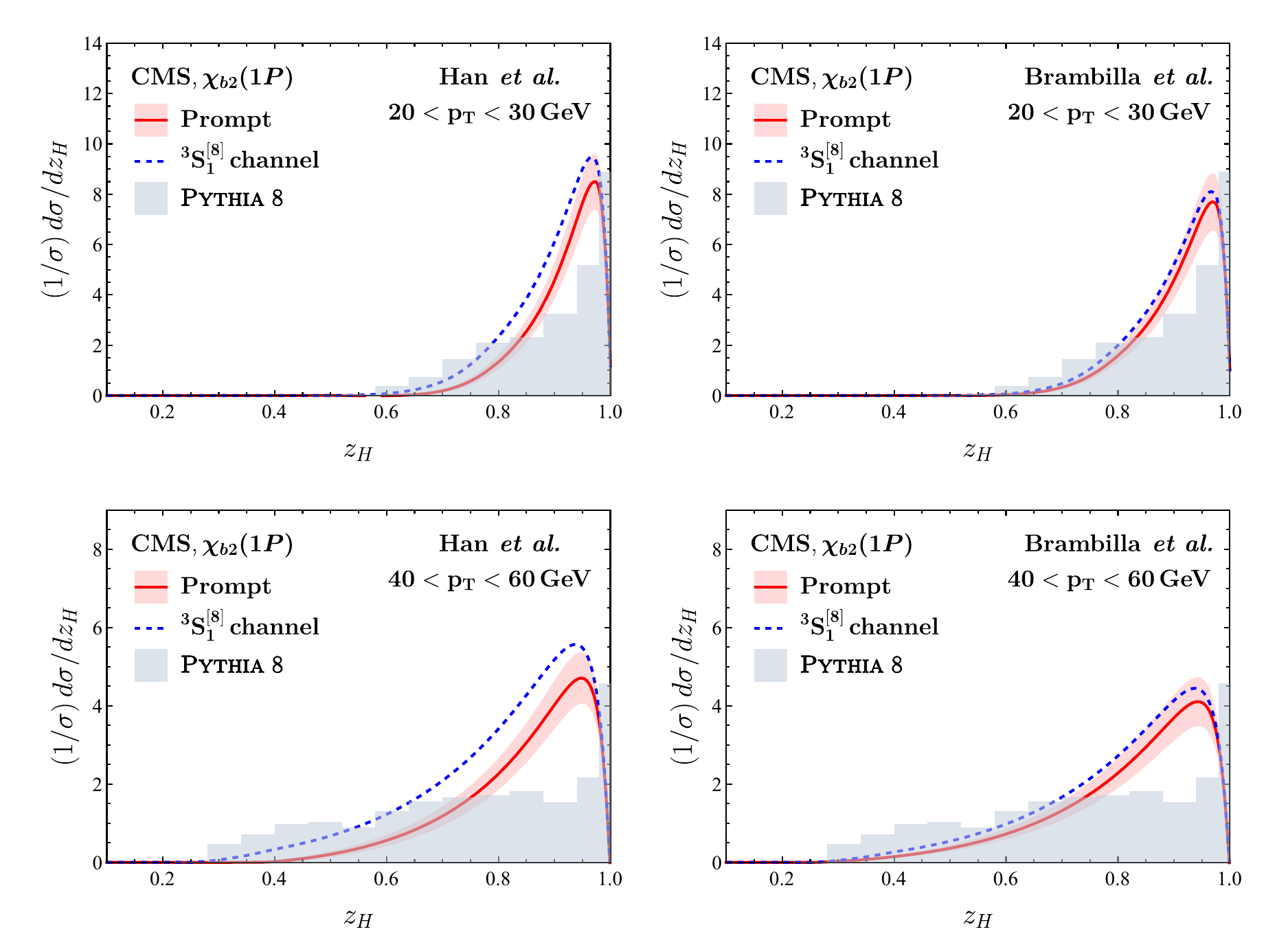}
    \end{adjustbox}
    \caption{
    Normalized $z_H$ distributions for inclusive $\chi_{b2}(1P)$ production with CMS kinematic selection.
    The upper and lower panels correspond to $20 < p_T < 30~\text{GeV}$ and $40 < p_T < 60~\text{GeV}$, respectively, while the left and right panels show the results for the two LDME sets.
    }
    \label{fig:chib21P_CMS_rep}
\end{figure}

\begin{figure}[h]
    \centering
    \begin{adjustbox}{center}
    \includegraphics[width=0.7\textwidth]{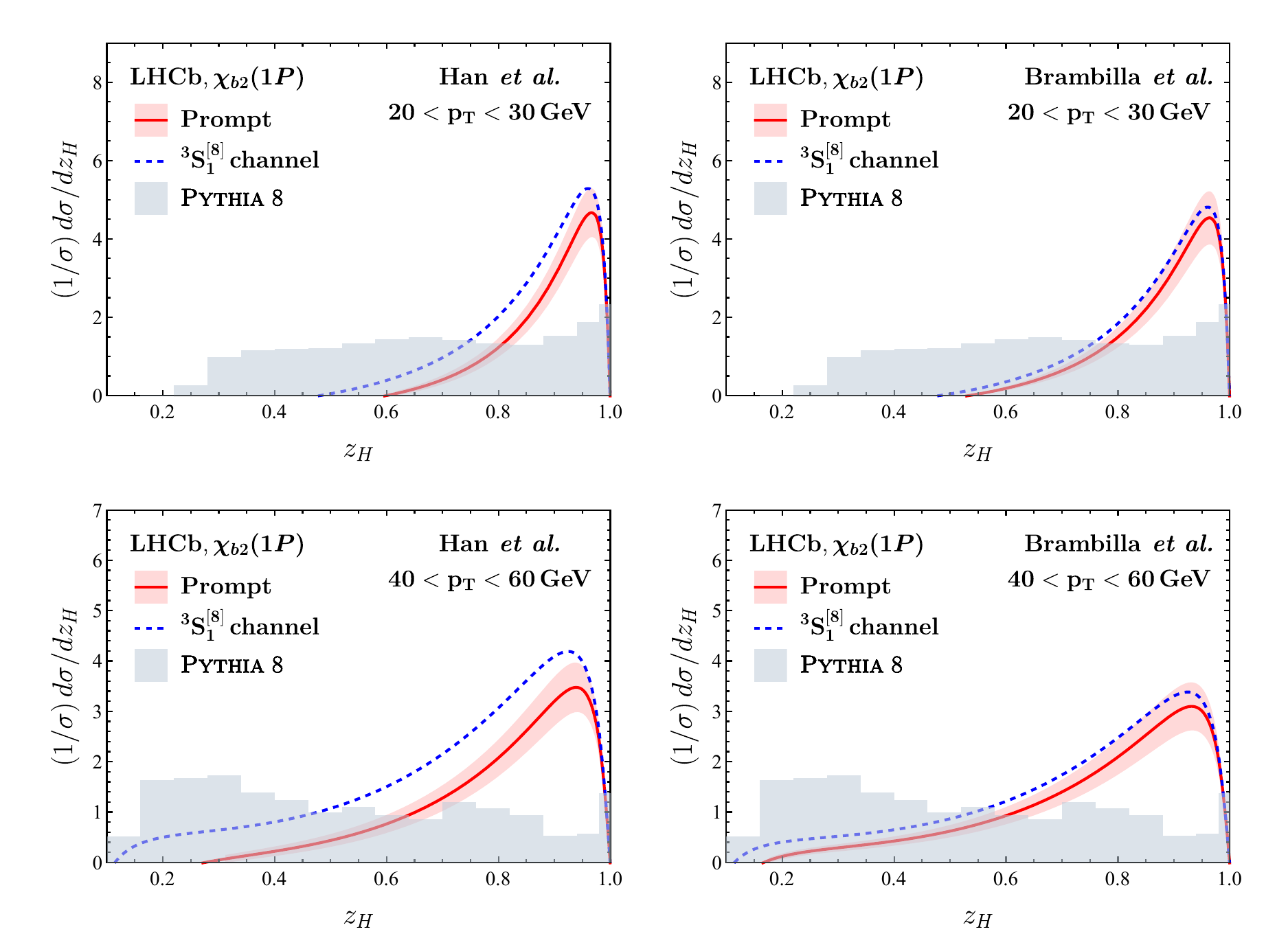}
    \end{adjustbox}
    \caption{
    Normalized $z_H$ distributions for inclusive $\chi_{b2}(1P)$ production with LHCb kinematic selection, using the same $p_T$ bins and LDME sets as in \fig{chib21P_CMS_rep}.
    }
    \label{fig:chib21P_LHCb_rep}
\end{figure}

We now turn to \emph{inclusive} $\chi_{bJ}(nP)$ production, which includes both direct production and feeddown contributions.
For direct production, in contrast to $\chi_{cJ}$ production, the ${}^3S_1^{[8]}$ contribution is generally much more pronounced for $\chi_{bJ}(nP)$, while the ${}^3P_J^{[1]}$ contribution provides only a relatively small correction to the direct distribution due to an enhanced LDME ratio of ${}^3S_1^{[8]}$ to ${}^3P_J^{[1]}$ as well as a larger SDC for the ${}^3S_1^{[8]}$ channel. Consequently, the shape of the direct distribution is largely governed by the ${}^3S_1^{[8]}$ channel.

This enhanced LDME ratio can be qualitatively understood from the pNRQCD description of the relevant LDMEs \cite{Brambilla:2021abf}. In the framework, the ratio divided by $m_Q^2$ is related to a universal gluonic correlator, which is independent of the heavy quarkonium states. The renormalization group evolution of this correlator from the charmonium to the bottomonium scale leads to an enhancement of this ratio, which helps explain the stronger ${}^3S_1^{[8]}$ contribution observed for bottomonium.

The feeddown contribution from higher bottomonium states is found to be small for both LDME sets. In particular, the contributions from higher $\Upsilon$ states are suppressed compared with direct $\chi_{bJ}$ production.

\fig{chib21P_CMS_rep} and \fig{chib21P_LHCb_rep} show the normalized $z_H$ distributions for $\chi_{b2}(1P)$ production at two representative $p_T$ bins under the CMS and LHCb kinematic selections, respectively. More results for the other $p_T$ bins and $J$ states are presented in the Appendices~\ref{app:zhCMS} and \ref{app:zhLHCb}.
The pronounced peak of the ${}^3S_1^{[8]}$ channel decreases with increasing $p_T$, but it remains dominant over the considered range for both LDME sets. 

A notable difference from the charmonium case is that the distributions obtained with the Han and Brambilla LDME sets are very similar despite differences in the underlying LDMEs. The relatively small ${}^3P_J^{[1]}$ contribution suppresses the impact of its $J$-dependent SDCs, resulting in nearly $J$-degenerate distributions. More generally, the $z_H$ distributions show universal behavior with remarkably weak dependence on the choice of LDME sets, the radial excitation ($n=1,2,3$), and the total angular momentum ($J=0,1,2$). Although the normalized $z_H$ distributions are largely insensitive to the LDME set, the ${}^3S_1^{[8]}$ LDME values differ by about a factor of two between the Han and Brambilla sets, leading to substantially different absolute cross sections.

The theoretical distributions obtained with the CMS~\cite{CMS:2026hnn} and LHCb~\cite{LHCb:2018yzj} kinematic selections are also very similar for bottomonium production. The different muon-acceptance requirements lead to a stronger suppression of the low-$z_H$ region for CMS than for LHCb, with the difference becoming less pronounced as the $p_T$ increases.

We also compare our predictions with the \textsc{Pythia}~8 results for $\chi_{b2}(1P)$ production. For the CMS selection, the \textsc{Pythia}~8 result in the low-$p_T$ bin ($20 < p_T < 30~\text{GeV}$) exhibits a pronounced peak in the highest-$z_H$ bin and decreases toward lower $z_H$, similar to the behavior observed for $\chi_{c2}$ production (see \fig{chic2_CMS_rep}). In the high-$p_T$ bin, the peak in theoretical results becomes less pronounced in both LDME sets, while the \textsc{Pythia}~8 distribution has quickly spread over wide $z_H$ region and shows a more gradual decrease toward lower $z_H$ than our prediction. For the LHCb selection, the \textsc{Pythia}~8 results exhibit relatively broad distributions for both $p_T$ bins. In the low-$p_T$ bin, the \textsc{Pythia}~8 distribution is considerably flatter than our prediction. In the high-$p_T$ bin, the \textsc{Pythia}~8 distribution increases slightly as $z_H$ decreases to around $0.2$, before dropping sharply in the lowest-$z_H$ bin. In contrast, our prediction exhibits a peak near the threshold and decreases toward lower $z_H$.

\section{Conclusion}
\label{sec:4}

In this paper, we have presented the first systematic theoretical study of $P$-wave charmonium and bottomonium production within jets at the LHC. Our calculations are formulated within the fragmenting jet function framework at NLO accuracy, with DGLAP evolution and threshold resummation included at LL accuracy. We have investigated the $z_H$ distributions using two representative sets of NRQCD long-distance matrix elements for both the charm and bottom sectors.

The underlying short-distance coefficients, which describe the perturbatively calculable contributions, exhibit distinct behavior for the two channels, ${}^3S_1^{[8]}$ and ${}^3P_J^{[1]}$. 
The ${}^3S_1^{[8]}$ channel appears at lower order in $\alpha_s$ and generally has a larger coefficient in magnitude than the ${}^3P_J^{[1]}$ channel, which starts at the next order. 
The ${}^3S_1^{[8]}$ coefficient exhibits a pronounced enhancement at large $z_H$ at relatively low $p_T$, which becomes broader and shifts toward smaller $z_H$ as the $p_T$ increases. In contrast, the ${}^3P_J^{[1]}$ coefficients are negative over most of the relevant kinematic region and exhibit nontrivial $J$ dependence.

For $P$-wave bottomonium, the normalized $z_H$ distributions exhibit a near-universal behavior: they are remarkably insensitive to the choice of long-distance matrix-element set, radial excitation $n$, and total angular momentum $J$. This universality is a consequence of the dominance of the ${}^3S_1^{[8]}$ channel, which makes the relative impact of the ${}^3P_J^{[1]}$ contribution small. The feeddown contributions from higher $\Upsilon$ states are also found to be small in the kinematic region considered. This contrasts with inclusive $\Upsilon$ production, where feeddown from $\chi_{bJ}(nP)$ states provides an important contribution to the observed yield. The relatively small feeddown contribution in $\chi_{bJ}(nP)$ production means that the observed $z_H$ distributions more directly reflect the direct production dynamics.

The charmonium sector shows a qualitatively different behavior. The smaller ${}^3S_1^{[8]}$ long-distance matrix element in the Brambilla set allows the ${}^3P_J^{[1]}$ contribution to have a visible impact on the normalized $z_H$ distributions. In particular, the negative ${}^3P_J^{[1]}$ contribution affects the overall $z_H$ shape, whereas the greater relative importance of the ${}^3S_1^{[8]}$ channel in the Bodwin set leads to a distribution that more closely follows the characteristic ${}^3S_1^{[8]}$ shape. The nontrivial $J$ dependence of the ${}^3P_J^{[1]}$ short-distance coefficients also leads to differences among the normalized distributions for $J=0,1,2$, particularly for the Brambilla set. Feeddown from $\psi(2S)$ is also more relevant for the Brambilla set, particularly for $\chi_{c0}$.

Our calculations also account for the effects of the experimental acceptance on the $z_H$ distributions. The muon selection requirements used by CMS and LHCb suppress the low-$z_H$ region, with the effect being more pronounced for CMS owing to its more restrictive muon kinematic requirements. The LHCb acceptance extends further toward smaller $z_H$ values. These effects should therefore be taken into account when comparing theoretical predictions with future measurements. We have also compared our predictions with \textsc{Pythia}~8 simulations for $\chi_{c2}$ and $\chi_{b2}(1P)$ production under the CMS and LHCb kinematic selections. While differences are observed near the threshold, possibly reflecting the different treatments of threshold effects, both approaches generally show a broadening of the $z_H$ distributions with increasing $p_T$.

Overall, the simpler leading-channel structure and relatively limited feeddown of $P$-wave quarkonium production make these states a useful complementary laboratory for studying quarkonium fragmentation and the underlying NRQCD production mechanisms. In particular, the universal behavior found for $P$-wave bottomonium provides a useful benchmark for the dominant color-octet fragmentation dynamics, while the stronger long-distance matrix-element dependence observed for charmonium offers sensitivity to the relative contributions of the color-singlet and color-octet channels. Our results therefore provide theoretical benchmarks for future experimental studies of $P$-wave quarkonium production within jets at the LHC.

\acknowledgments 
The work of D.~Kang was supported by the faculty research fund of Sejong University in 2026.
The work of B.~Hong, B.~Putra, and S.~Lee was supported by the National Research Foundation of Korea (NRF) grant funded by the Korea government (MSIT) (No.~RS-2008-NR007227).


\appendix

\section{Feeddown LDME}
\label{app:feeddownLDME}

\begin{table}[thb]
	\centering
	\begin{tabular}{cccccc}
		\hline
        \multirow{2}{*}{\textbf{State}} 
		& \multirow{2}{*}{\phantom{Empty}} 
		& $\langle \mathcal{O}(^{3}S^{[1]}_{1}) \rangle$
		& $\langle \mathcal{O}(^{3}S^{[8]}_{1}) \rangle$ 
		& $\langle \mathcal{O}(^{1}S^{[8]}_{0}) \rangle$
		& $\langle \mathcal{O}(^{3}P^{[8]}_{0}) \rangle / m_c^2$ \\ 
		&
		& $\times\text{GeV}^3$
		& $\times10^{-2}~\text{GeV}^3$ 
		& $\times10^{-2}~\text{GeV}^3$
		& $\times10^{-2}~\text{GeV}^3$ \\ \hline\hline
		\multirow{2}{*}{$\psi(2S)$}
		& Brambilla \textit{et al.}\cite{Brambilla:2022ayc}
		& $0.71\pm0.21$ & $0.84\pm0.25$ & $-0.37\pm1.92$ & $1.55\pm0.49$ \\ \cline{2-6}
		
		& Bodwin \textit{et al.}\cite{Bodwin:2015iua}
		& $0.76$ & $-0.157\pm0.280$ & $3.14\pm0.79$ & $-0.114\pm0.121$ \\ \hline\hline
	\end{tabular}
    \caption{Representative $\psi(2S)$ LDME sets used in the calculation, provided in units of $\text{GeV}^3$ for ${}^3S_1^{[1]}$ and $10^{-2}\,\text{GeV}^3$ for the other channels, with $m_c=1.5$~GeV.}
	\label{tab:psi2S_ME}
\end{table}

\begin{table}[thb]
	\centering
	\begin{tabular}{cccccc}
		\hline
        \multirow{2}{*}{\textbf{State}} 
		& \multirow{2}{*}{\phantom{Empty}} 
		& $\langle \mathcal{O}(^{3}S^{[1]}_{1}) \rangle$
		& $\langle \mathcal{O}(^{3}S^{[8]}_{1}) \rangle$ 
		& $\langle \mathcal{O}(^{1}S^{[8]}_{0}) \rangle$
		& $\langle \mathcal{O}(^{3}P^{[8]}_{0}) \rangle / m_b^2$ \\ 
		&
		& $\times\text{GeV}^3$
		& $\times10^{-2}~\text{GeV}^3$ 
		& $\times10^{-2}~\text{GeV}^3$
		& $\times10^{-2}~\text{GeV}^3$ \\ \hline\hline
		
		\multirow{2}{*}{$\Upsilon(2S)$}
		& Han \textit{et al.}\cite{Han:2014kxa}
		& $4.63$ & $1.08\pm0.20$ & $6.07\pm1.08$ & $0$ \\ \cline{2-6}
		
		& Brambilla \textit{et al.}\cite{Brambilla:2022ayc}
		& $4.96\pm0.50$ & $1.52\pm0.47$ & $-0.20\pm1.04$ & $1.08\pm0.35$ \\ \hline\hline
		
		\multirow{2}{*}{$\Upsilon(3S)$}		
		&\multirow{2}{*}{Same as $2S$}
		& $3.54$ & $0.83\pm0.02$ & $2.83\pm0.07$ & $0$ \\ \cline{3-6}
		
		& 
		& $3.83\pm0.38$ & $1.17\pm0.37$ & $-0.16\pm0.81$ & $0.84\pm0.27$ \\ \hline\hline
	\end{tabular}
    \caption{Representative $\Upsilon(2S)$ and $\Upsilon(3S)$ LDME sets used in the calculation, provided in units of $\text{GeV}^3$ for ${}^3S_1^{[1]}$ and $10^{-2}\,\text{GeV}^3$ for the other channels, with $m_b=4.75$~GeV.}
	\label{tab:Upsilon_ME}
\end{table}

The feeddown contributions from higher charmonium and bottomonium states are evaluated using the LDME sets summarized in \tab{psi2S_ME} and \tab{Upsilon_ME}. For $\psi(2S)$, we consider the Brambilla and Bodwin LDME sets, while for $\Upsilon(2S)$ and $\Upsilon(3S)$ we use the Han and Brambilla sets. For the Han set, we adopt the representative solution obtained by maximizing $\langle O(^1S_0^{[8]})\rangle$ under the positivity constraint. The branching fractions for the established states are taken from the Particle Data Group~\cite{ParticleDataGroup:2024cfk}, while those for the $\chi_{bJ}(3P)$ transitions are taken from the predictions of Ref.~\cite{Han:2014kxa}.

\section{Muon acceptance}
\label{app:mucut}

$P$-wave quarkonia are experimentally identified through radiative decays into $S$-wave quarkonia followed by the decay of the daughter quarkonium into a muon pair,
$\chi_{QJ}\to H_S+\gamma\to\mu^+\mu^-+\gamma$,
where $H_S=J/\psi$ for charmonium and $H_S=\Upsilon(mS)$ for bottomonium.
Because the daughter $S$-wave quarkonium carries only a fraction of the parent $P$-wave quarkonium momentum after the radiative decay, the muon kinematics and hence the muon acceptance differ from those for directly produced $S$-wave quarkonia, even when the same muon selection requirements are applied.

In our analysis, we apply the same muon-pair selection requirements used in the corresponding experimental measurements of $S$-wave quarkonia.
In principle, the acceptance depends on the specific $P$-wave state and decay channel, and therefore on $J$, the radial excitation $n$, and the daughter-state quantum number $m$, through the corresponding decay kinematics and particle masses. We have checked these dependences in our simulation and find that the resulting differences in the acceptance are small. We therefore use the $J=1$ and $n=m$ configurations as representative cases, which also correspond to decay channels with relatively large branching fractions, for the acceptance corrections presented below.

\begin{figure}[t]
\centering

\begin{subfigure}{0.42\textwidth}
\centering
\includegraphics[width=\textwidth]{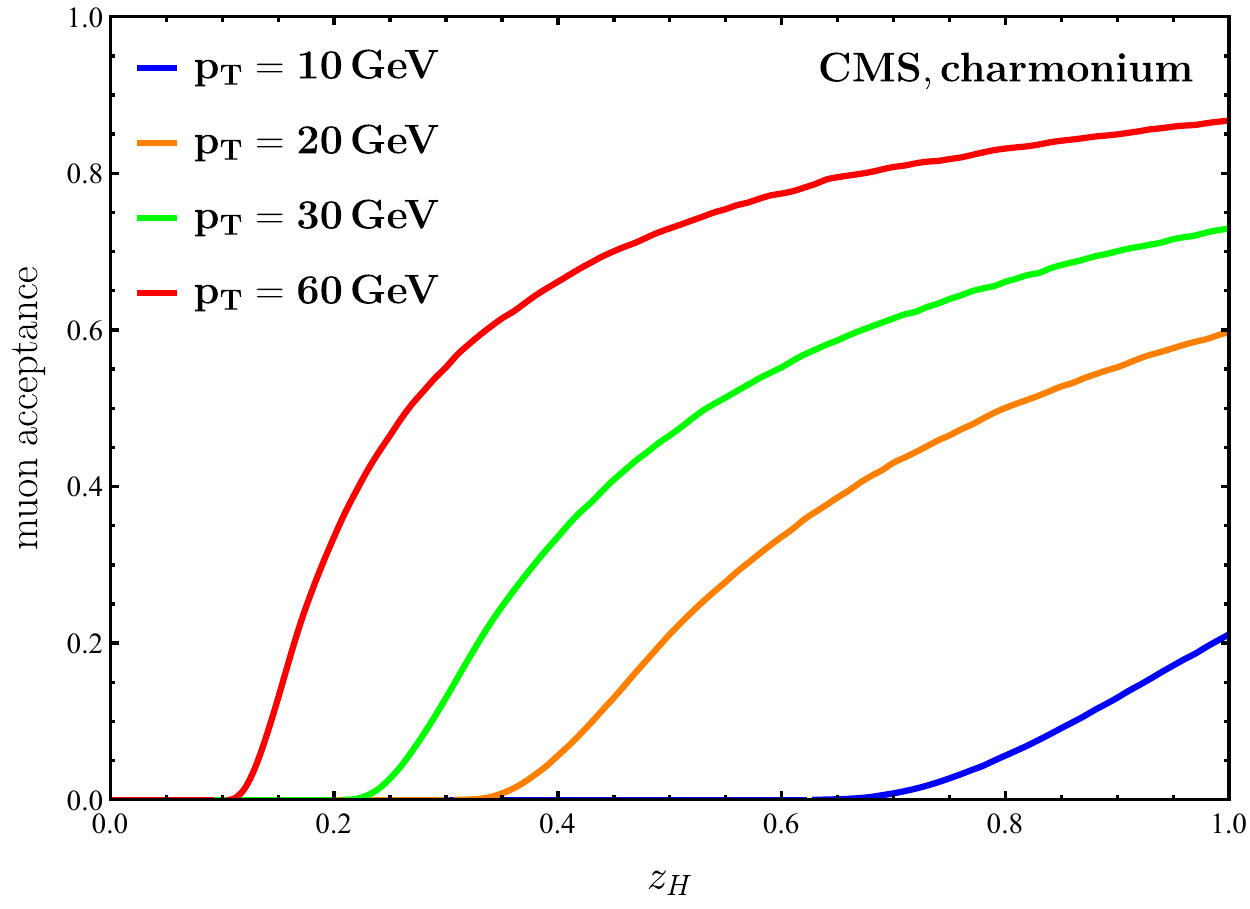}
\caption{CMS, charmonium}
\label{fig}
\end{subfigure}
\hspace{0.04\textwidth}
\begin{subfigure}{0.42\textwidth}
\centering
\includegraphics[width=\textwidth]{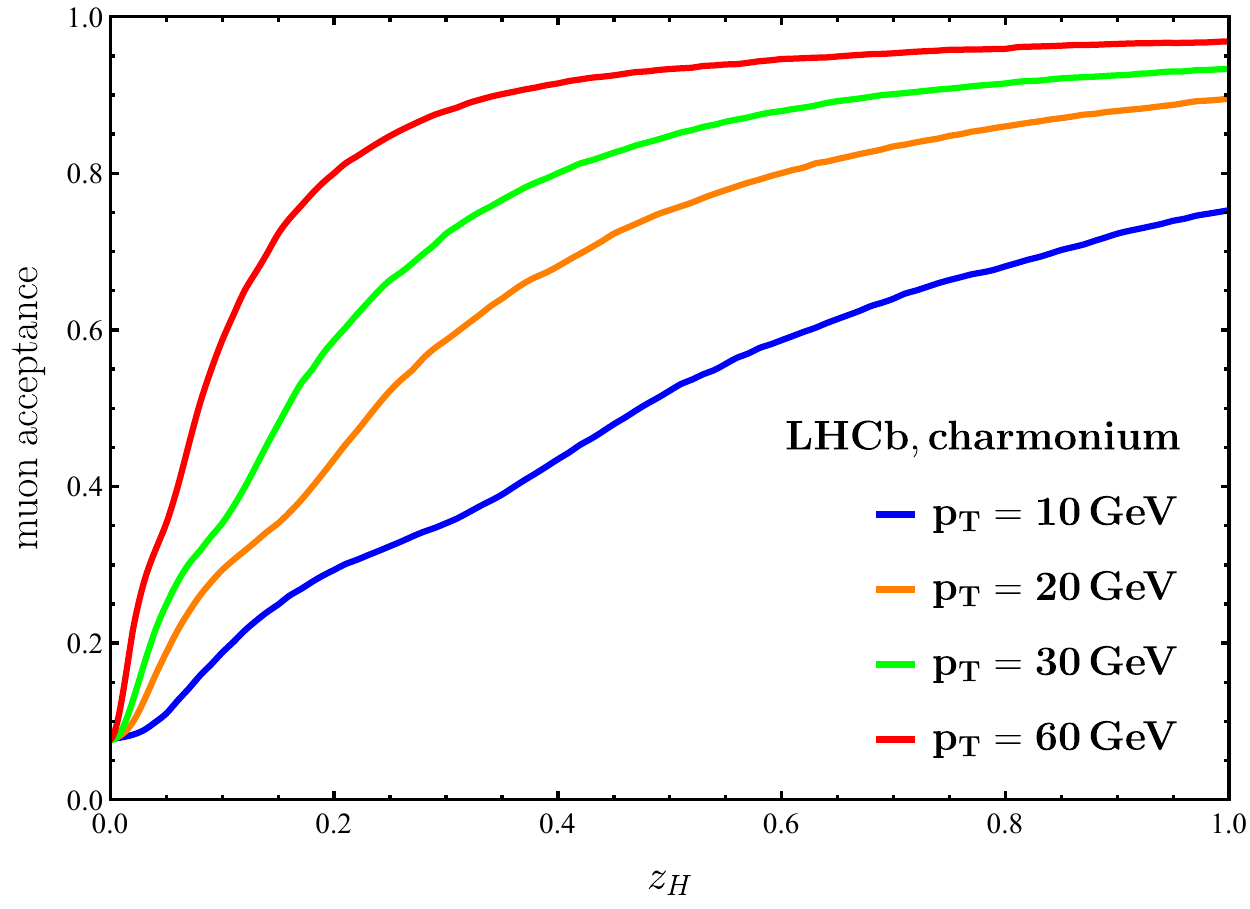}
\caption{LHCb, charmonium}
\label{fig}
\end{subfigure}

\vspace{0.4cm}

\begin{subfigure}{0.42\textwidth}
\centering
\includegraphics[width=\textwidth]{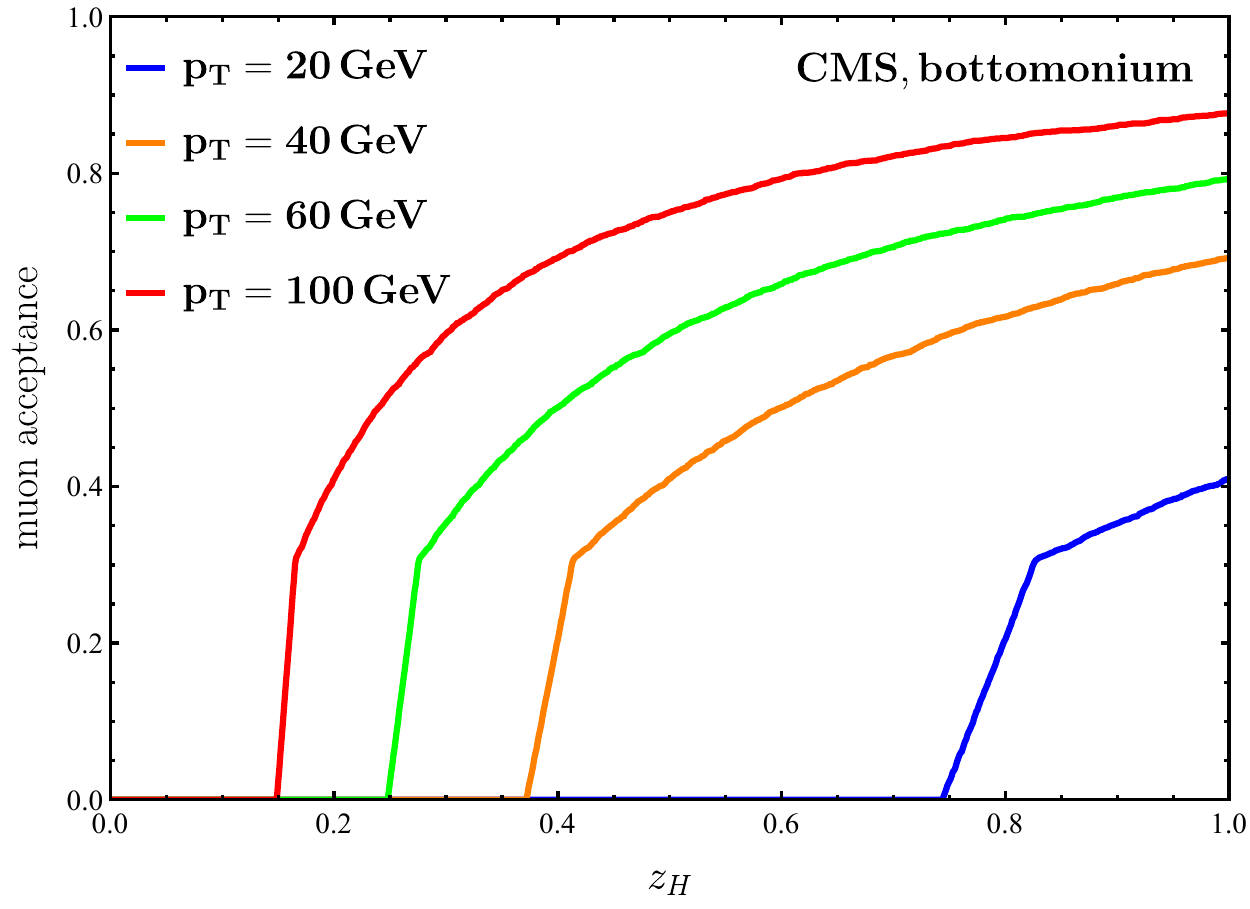}
\caption{CMS, bottomonium}
\label{fig}
\end{subfigure}
\hspace{0.04\textwidth}
\begin{subfigure}{0.42\textwidth}
\centering
\includegraphics[width=\textwidth]{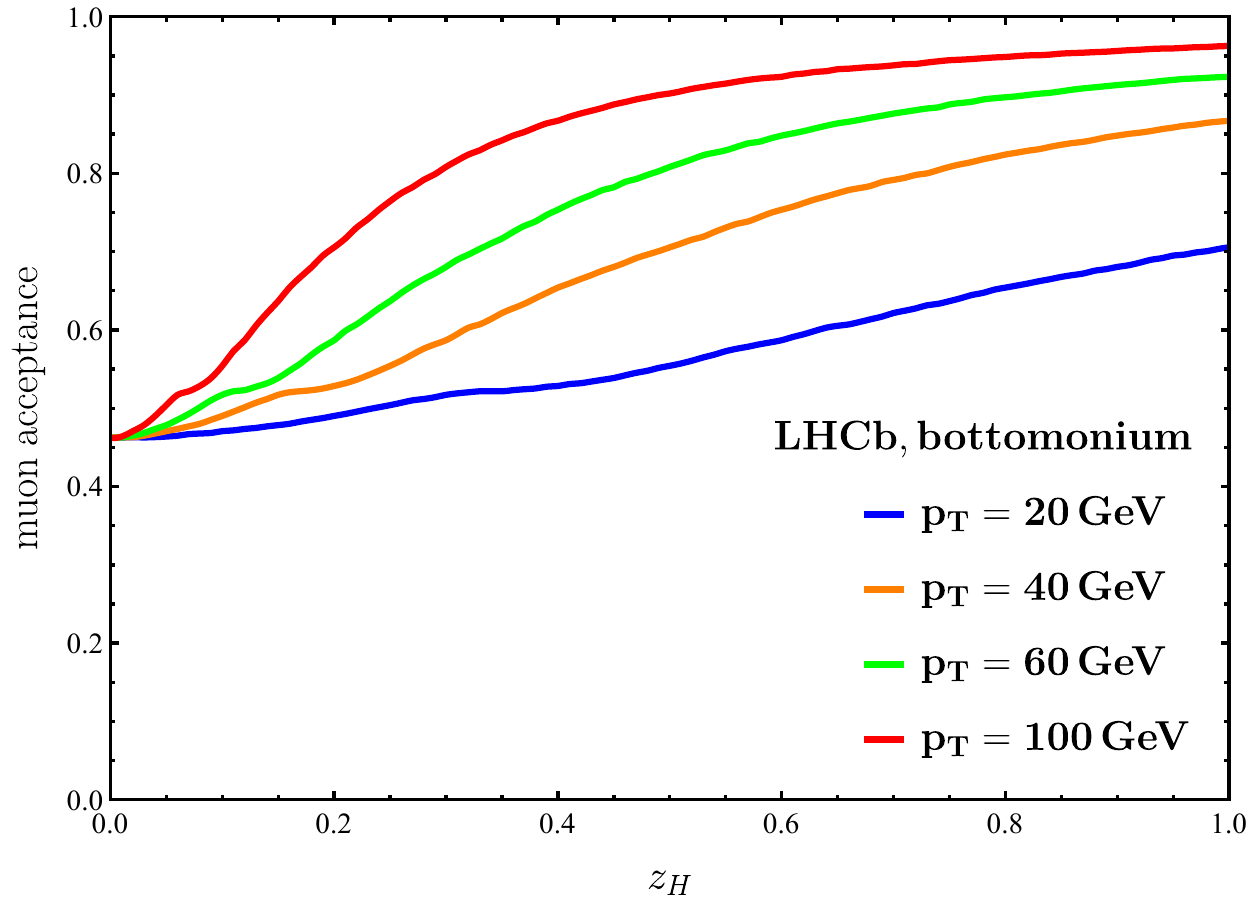}
\caption{LHCb, bottomonium}
\label{fig}
\end{subfigure}

\caption{
Muon acceptance effects on the $z_H$ distributions for CMS (left column) and LHCb (right column) kinematic selections. The top row shows results for charmonium ($\chi_{c1}$) and the bottom row for bottomonium ($\chi_{b1}(1P)$).
}
\label{fig:mucut}
\end{figure}

The impact of these muon acceptance requirements on the $z_H$ distributions is illustrated in \fig{mucut}. The left and right columns show the results for the CMS and LHCb selections, respectively, while the top and bottom rows correspond to charmonium ($\chi_{c1}$) and bottomonium ($\chi_{b1}(1P)$), respectively. The results are shown for representative $p_T$ bins, with the jet pseudorapidity fixed to $\eta=0$ for CMS and $\eta=2.5$ for LHCb.

The muon selection requirements used in our analysis are summarized as follows:

\begin{itemize}
\item \textbf{CMS, charmonium}~\cite{CMS:2021znk,CMS:2021mzx}: each muon satisfies
\begin{equation}
\begin{cases}
p_T(\mu)>3.5~\text{GeV}, & |\eta(\mu)|<1.2,\\
p_T(\mu)>(5.47-1.89|\eta(\mu)|)~\text{GeV}, & 1.2<|\eta(\mu)|<2.1,\\
p_T(\mu)>1.5~\text{GeV}, & 2.1<|\eta(\mu)|<2.4.
\end{cases}
\end{equation}

\item \textbf{LHCb, charmonium}~\cite{LHCb:2017llq}: both muons satisfy
\begin{equation}
2.0<\eta(\mu)<4.5,\qquad
p_T(\mu)>0.5~\text{GeV},
\end{equation}
together with
\begin{equation}
p(\mu)>5~\text{GeV},\qquad
\sqrt{p_T(\mu^+)p_T(\mu^-)}>1.5~\text{GeV}.
\end{equation}

\item \textbf{CMS, bottomonium}~\cite{CMS:2026hnn}: each muon satisfies
\begin{equation}
p_T(\mu)>6~\text{GeV},\qquad
|\eta(\mu)|<1.4,
\end{equation}
together with
\begin{equation}
p_T^{\mu^+\mu^-}>15~\text{GeV},\qquad
|y^{\mu^+\mu^-}|<1.2.
\end{equation}

\item \textbf{LHCb, bottomonium}~\cite{LHCb:2018yzj}: both muons satisfy
\begin{equation}
p_T(\mu)>1~\text{GeV},\qquad
p(\mu)>10~\text{GeV},\qquad
2<\eta(\mu)<4.5,
\end{equation}
together with
\begin{equation}
p_T(\mu^+)p_T(\mu^-)>(1.3~\text{GeV})^2.
\end{equation}
\end{itemize}

Compared with the corresponding muon acceptance for directly produced $S$-wave quarkonia, the $P$-wave decay chain leads to a somewhat different acceptance, particularly at relatively low $p_T$. This difference becomes increasingly small as the $p_T$ increases, and the $P$-wave acceptance is generally slightly smaller than the corresponding $S$-wave acceptance. The effect is somewhat more visible for the CMS selection, while the difference is relatively small for LHCb.

Both selections induce a suppression at low $z_H$, which originates from the muon reconstruction requirements rather than from the underlying quarkonium production dynamics. Since the muon cuts impose constraints on the momentum of the daughter $S$-wave quarkonium, they translate into an effective lower bound on $z_H=p_T^H/p_T$ that depends on the $p_T$.

The acceptance suppression is therefore weaker at higher $p_T$, where the same muon requirements correspond to smaller values of $z_H$. The difference between the CMS and LHCb selections is also reflected in \fig{mucut}: the LHCb selection allows the distribution to extend further into the low-$z_H$ region, while the CMS selection produces a more pronounced suppression there.

This purely kinematic effect should be distinguished from the low-$z_H$ behavior arising from fragmentation evolution and feeddown contributions discussed in the main text.

\section{$\chi_{cJ}$ and $\chi_{bJ}(nP)$ with CMS selections}
\label{app:zhCMS}

This section presents additional $z_H$ distributions for $\chi_{cJ}$ and $\chi_{bJ}(nP)$ production with the CMS kinematic selections. For bottomonium, the distributions for $\chi_{bJ}(3P)$ are not shown, as they are nearly identical to those for $\chi_{bJ}(2P)$, with only minor differences arising from the corresponding feeddown contributions. The $J$ dependence of the normalized bottomonium distributions is also very weak; therefore, only representative $J$ values are shown.

\begin{figure}[h]
    \centering
    \includegraphics[width=0.73\textwidth]{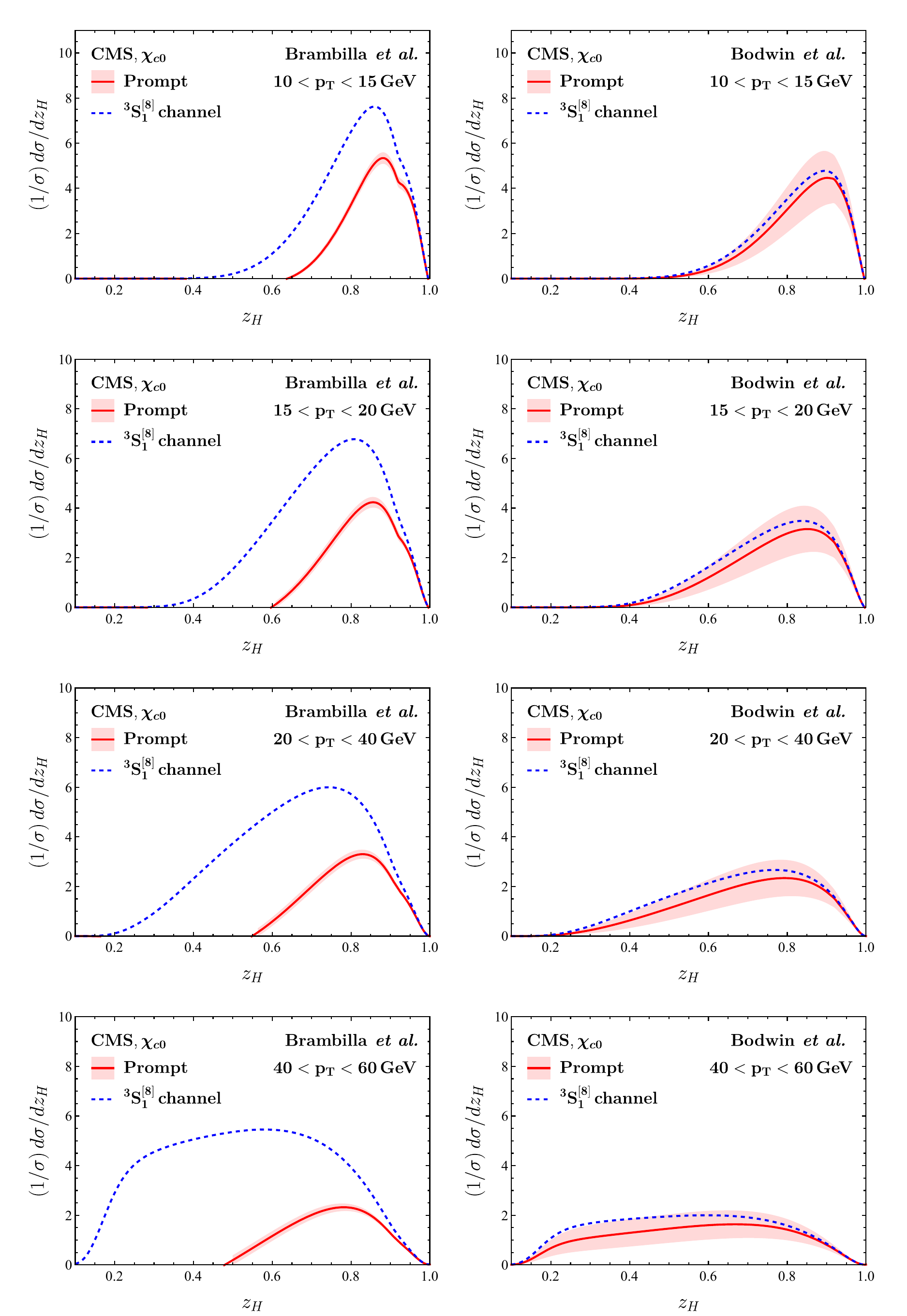}
    \caption{
        Normalized $z_H$ distributions for $\chi_{c0}$ production with the CMS kinematic selection.
        The left and right columns correspond to the Brambilla and Bodwin LDME sets, respectively, while the four rows correspond, from top to bottom, to the $p_T$ bins $10$--$15$, $15$--$20$, $20$--$40$, and $40$--$60~\text{GeV}$.
    }
    \label{fig:chic0_CMS}
\end{figure}

\begin{figure}[h]
    \centering
    \includegraphics[width=0.73\textwidth]{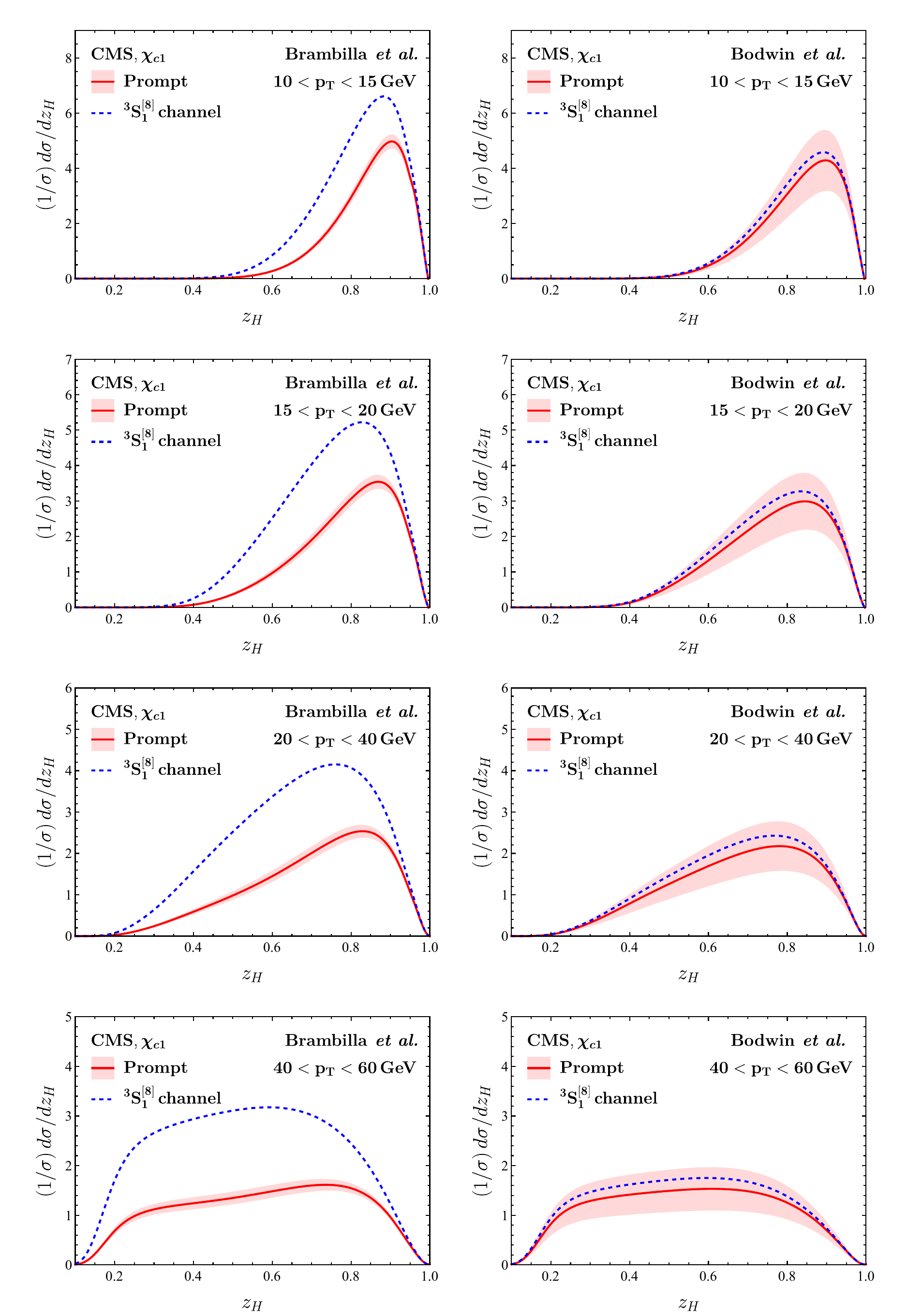}
    \caption{
        Normalized $z_H$ distribution for $\chi_{c1}$ production with the CMS kinematic selection.
        The left and right columns correspond to the Brambilla and Bodwin LDME sets, respectively, and the rows correspond to increasing $p_T$ bins.
    }
    \label{fig:chic1_CMS}
\end{figure}

\begin{figure}[h]
    \centering
    \includegraphics[width=0.73\textwidth]{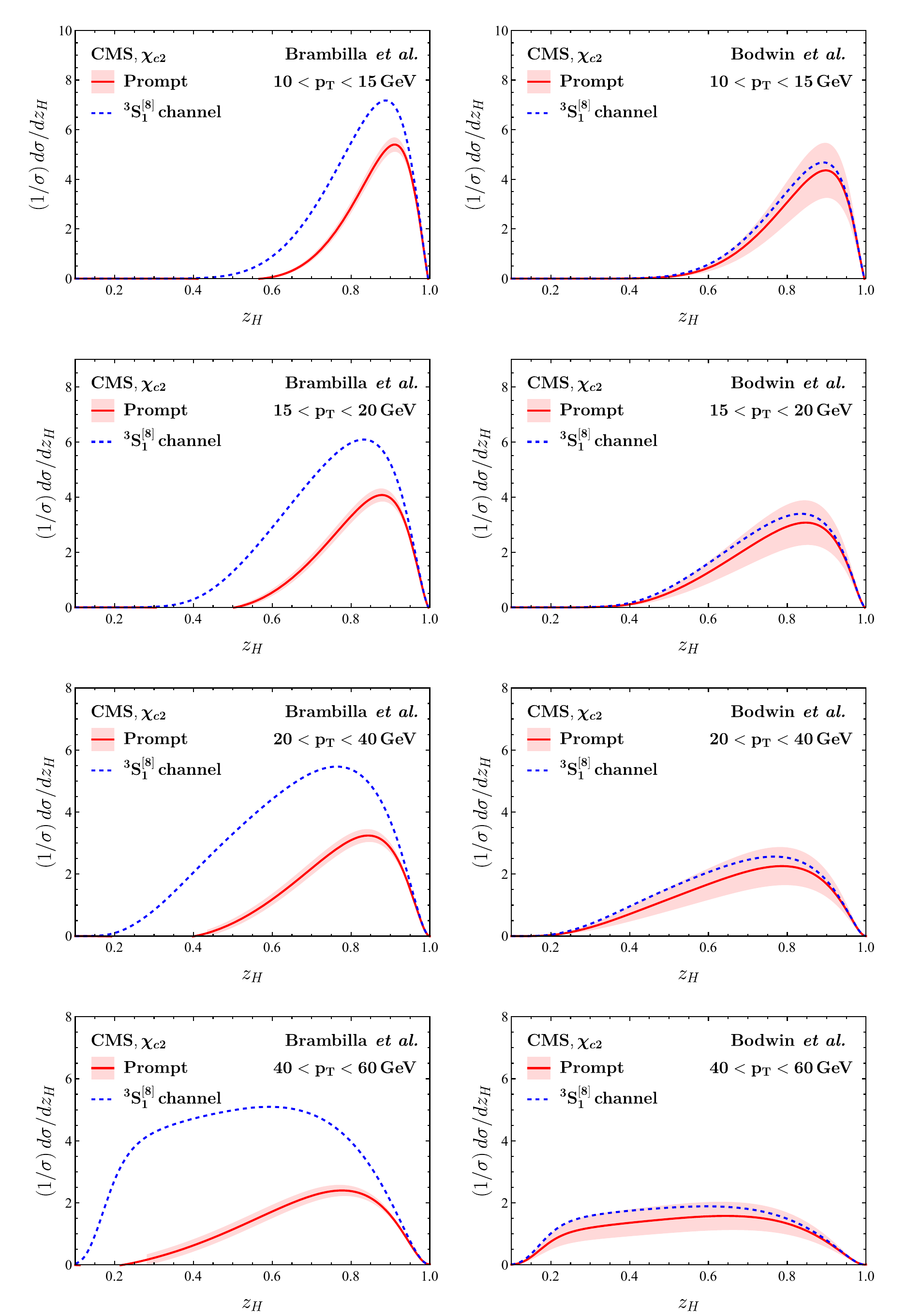}
    \caption{
        Normalized $z_H$ distribution for $\chi_{c2}$ production with the CMS kinematic selection.
        The left and right columns correspond to the Brambilla and Bodwin LDME sets, respectively, and the rows correspond to increasing $p_T$ bins.
    }
    \label{fig:chic2_CMS}
\end{figure}

\begin{figure}[thb]
    \centering
    \includegraphics[width=0.73\textwidth]{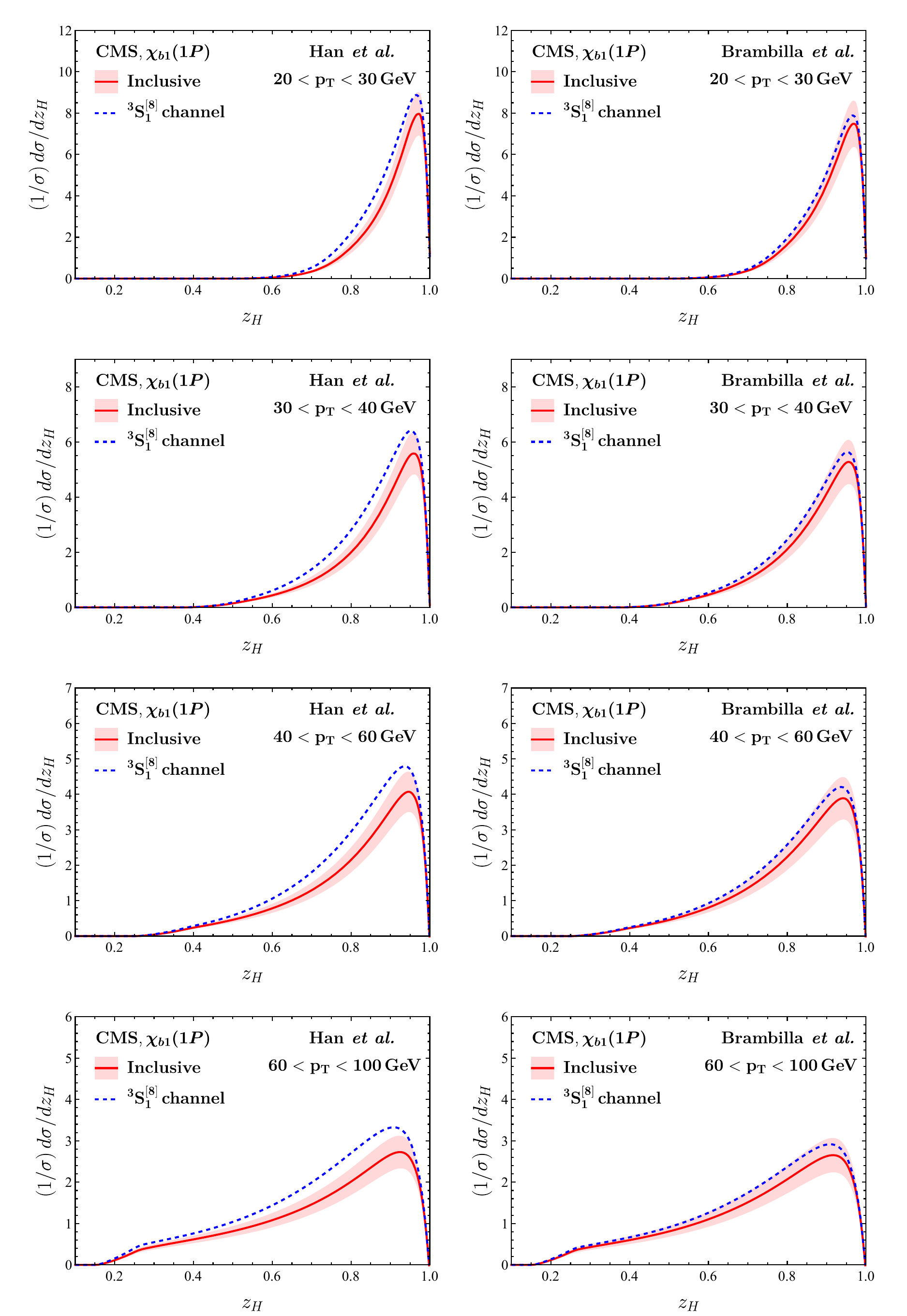}
    \caption{
        Normalized $z_H$ distribution for $\chi_{b1}(1P)$ production with the CMS kinematic selection.
        The left and right columns correspond to the Han and Brambilla LDME sets, respectively, while the four rows correspond, from top to bottom, to the $p_T$ bins $20$--$30$, $30$--$40$, $40$--$60$, and $60$--$100~\text{GeV}$.
    }
    \label{fig:chib11P_CMS}
\end{figure}

\begin{figure}[h]
    \centering
    \includegraphics[width=0.73\textwidth]{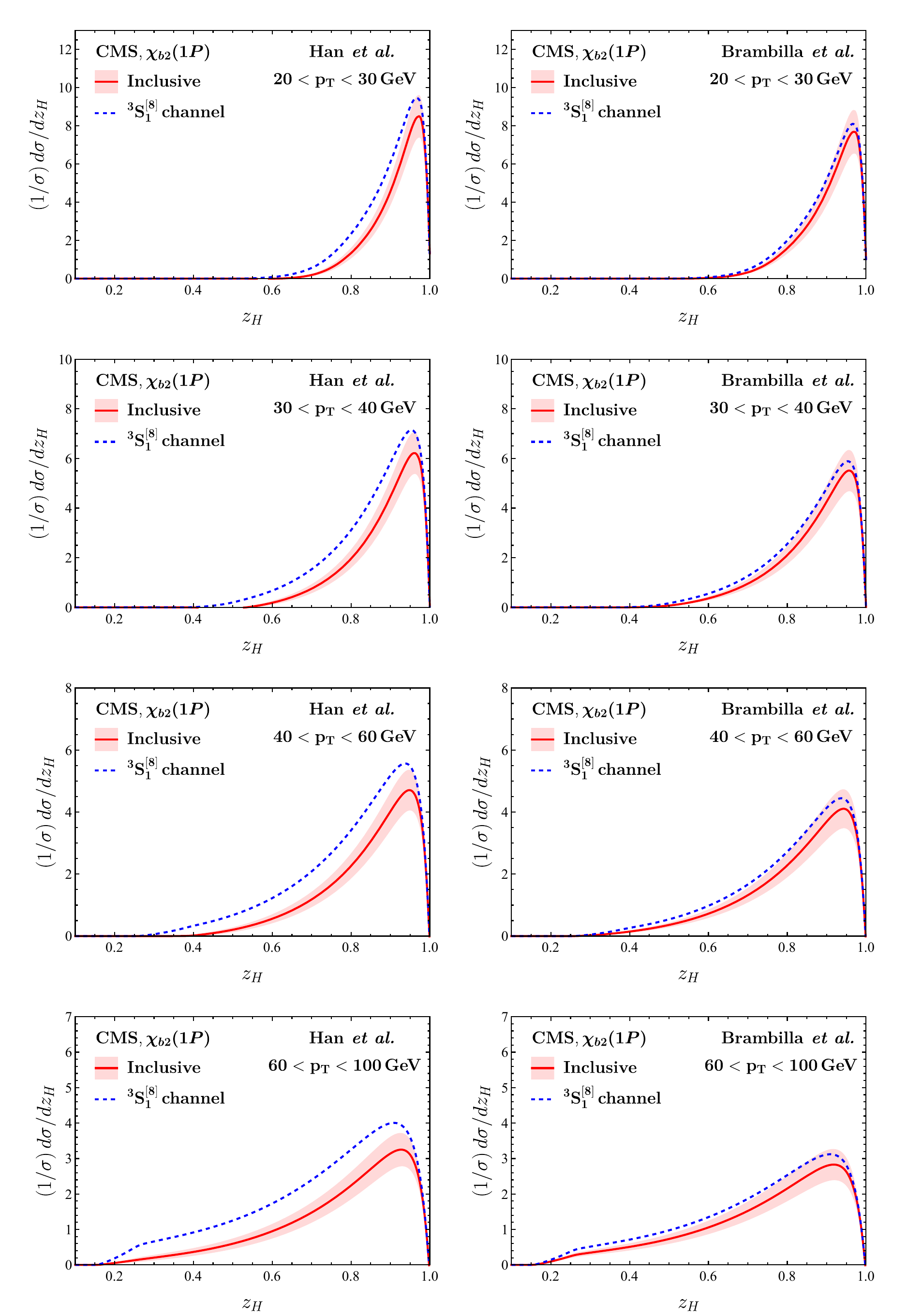}
    \caption{
        Normalized $z_H$ distribution for $\chi_{b2}(1P)$ production with the CMS kinematic selection.
        The left and right columns correspond to the Han and Brambilla LDME sets, respectively, and the rows correspond to increasing $p_T$ bins.
    }
    \label{fig:chib21P_CMS}
\end{figure}

\begin{figure}[h]
    \centering
    \includegraphics[width=0.73\textwidth]{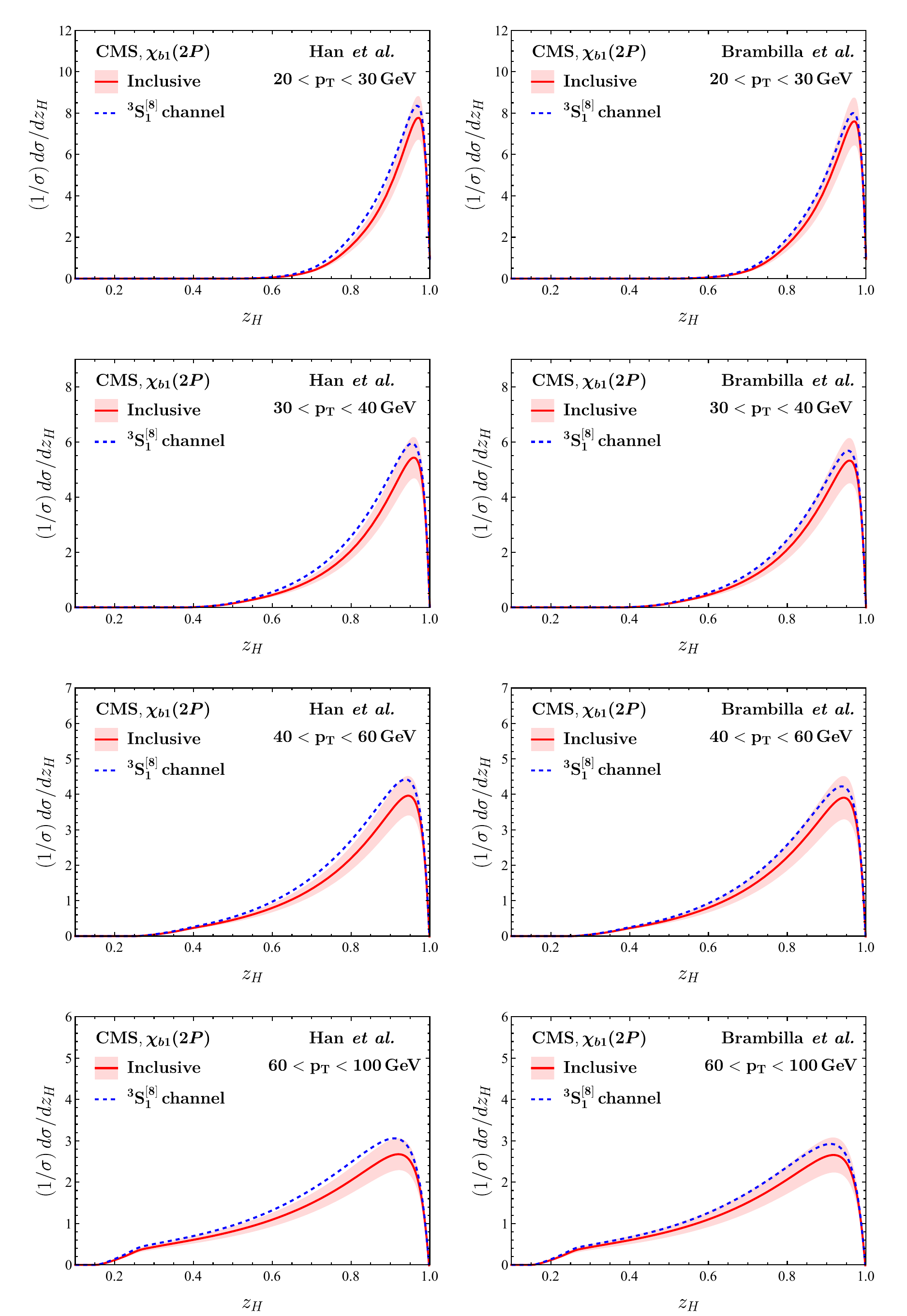}
    \caption{
        Normalized $z_H$ distribution for $\chi_{b1}(2P)$ production with the CMS kinematic selection.
        The left and right columns correspond to the Han and Brambilla LDME sets, respectively, and the rows correspond to increasing $p_T$ bins.
    }
    \label{fig:chib12P_CMS}
\end{figure}

\begin{figure}[h]
    \centering
    \includegraphics[width=0.73\textwidth]{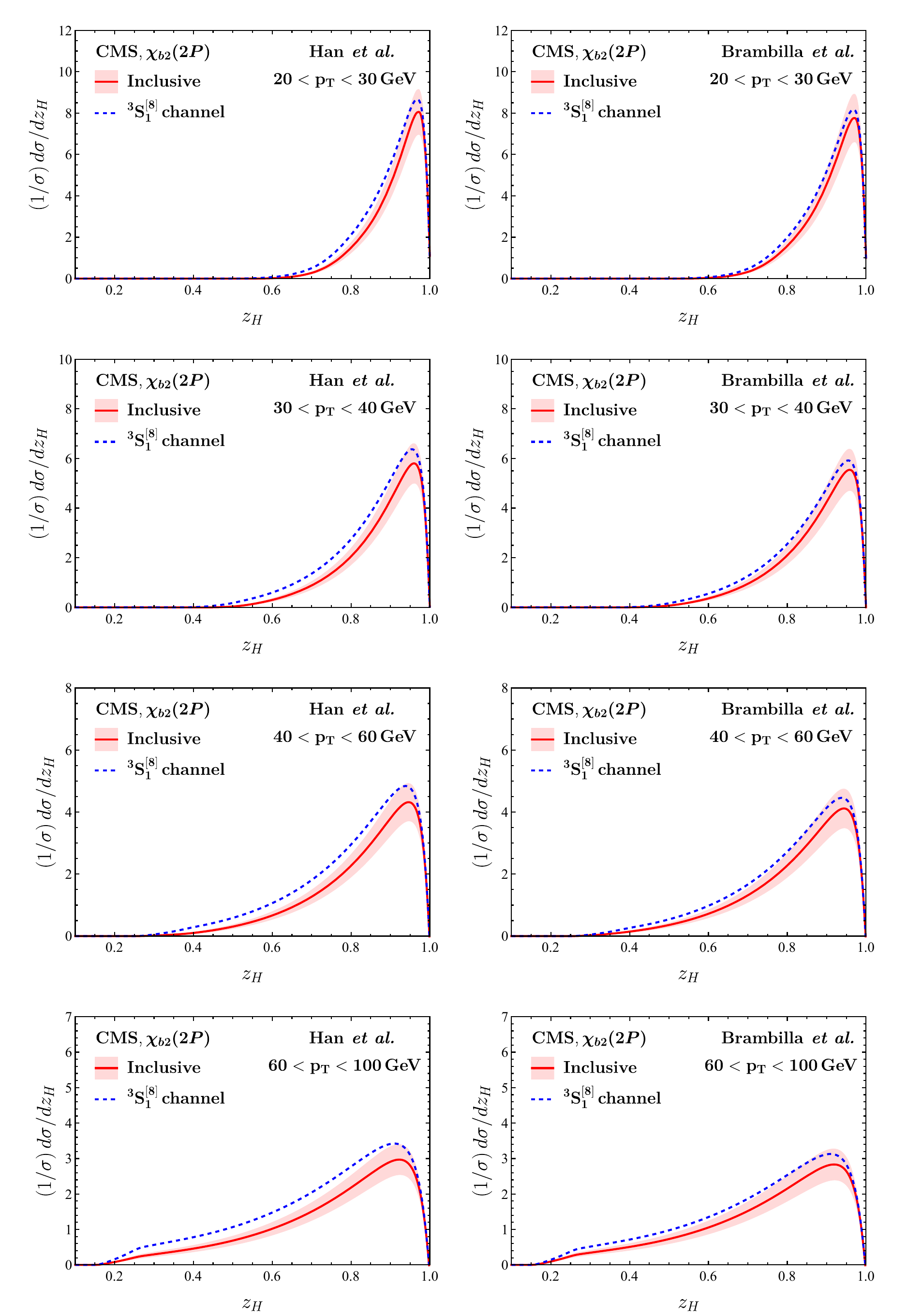}
    \caption{
        Normalized $z_H$ distribution for $\chi_{b2}(2P)$ production with the CMS kinematic selection.
        The left and right columns correspond to the Han and Brambilla LDME sets, respectively, and the rows correspond to increasing $p_T$ bins.
    }
    \label{fig:chib22P_CMS}
\end{figure}

\FloatBarrier

\section{$\chi_{cJ}$ and $\chi_{bJ}(nP)$ with LHCb selections}
\label{app:zhLHCb}

This section presents additional $z_H$ distributions for $\chi_{cJ}$ and $\chi_{bJ}(nP)$ production with the LHCb kinematic selections. For bottomonium, the distributions for $\chi_{bJ}(3P)$ are not shown, as they are nearly identical to those for $\chi_{bJ}(2P)$, with only minor differences arising from the corresponding feeddown contributions. The $J$ dependence is also very weak; therefore, only representative $J$ values are shown.

\begin{figure}[h]
    \centering
    \includegraphics[width=0.73\textwidth]{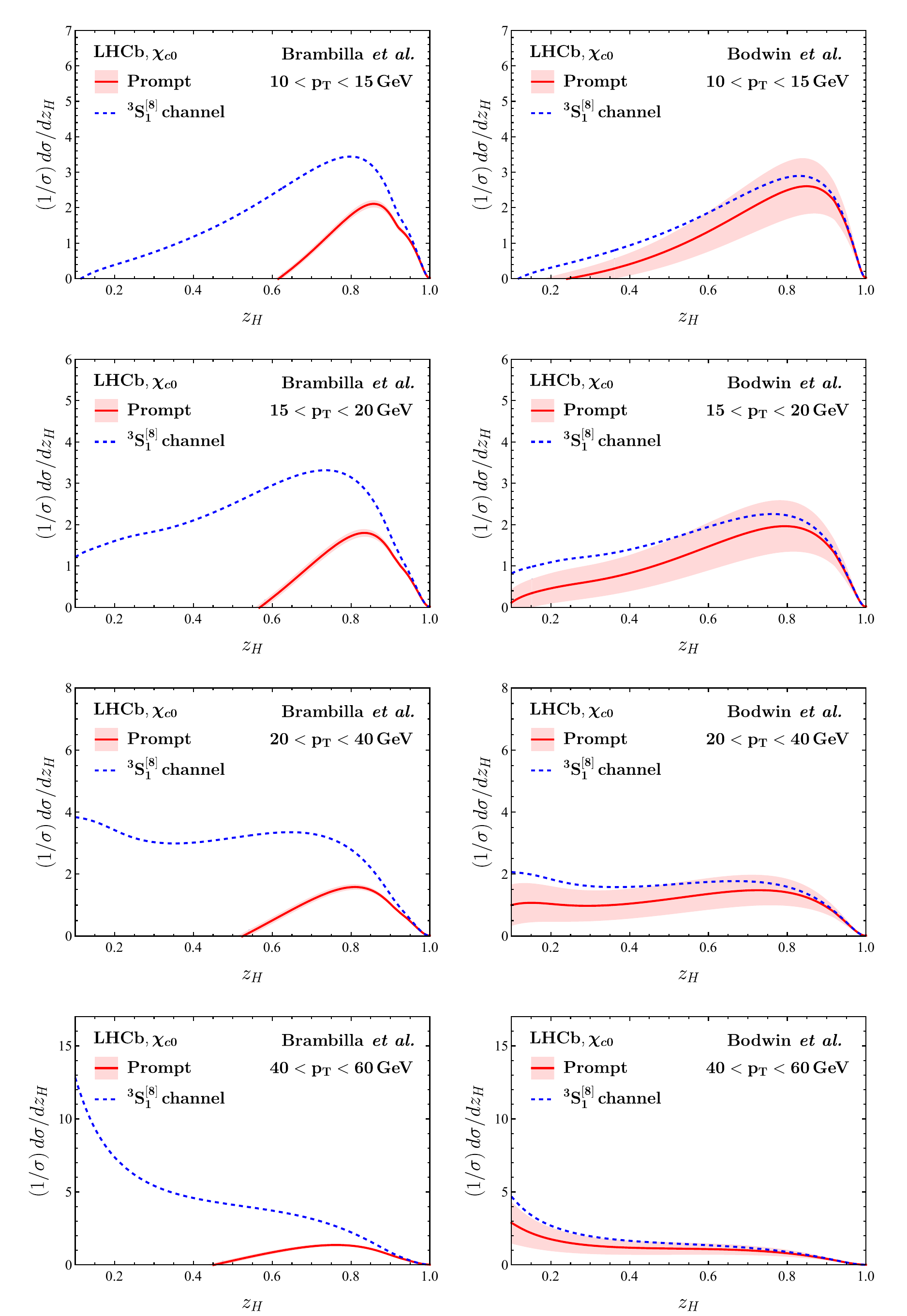}
    \caption{
        Normalized $z_H$ distribution for $\chi_{c0}$ production with the LHCb kinematic selection.
        The left and right columns correspond to the Brambilla and Bodwin LDME sets, respectively, while the four rows correspond, from top to bottom, to the $p_T$ bins $10$--$15$, $15$--$20$, $20$--$40$, and $40$--$60~\text{GeV}$.
    }
    \label{fig:chic0_LHCb}
\end{figure}

\begin{figure}[h]
    \centering
    \includegraphics[width=0.73\textwidth]{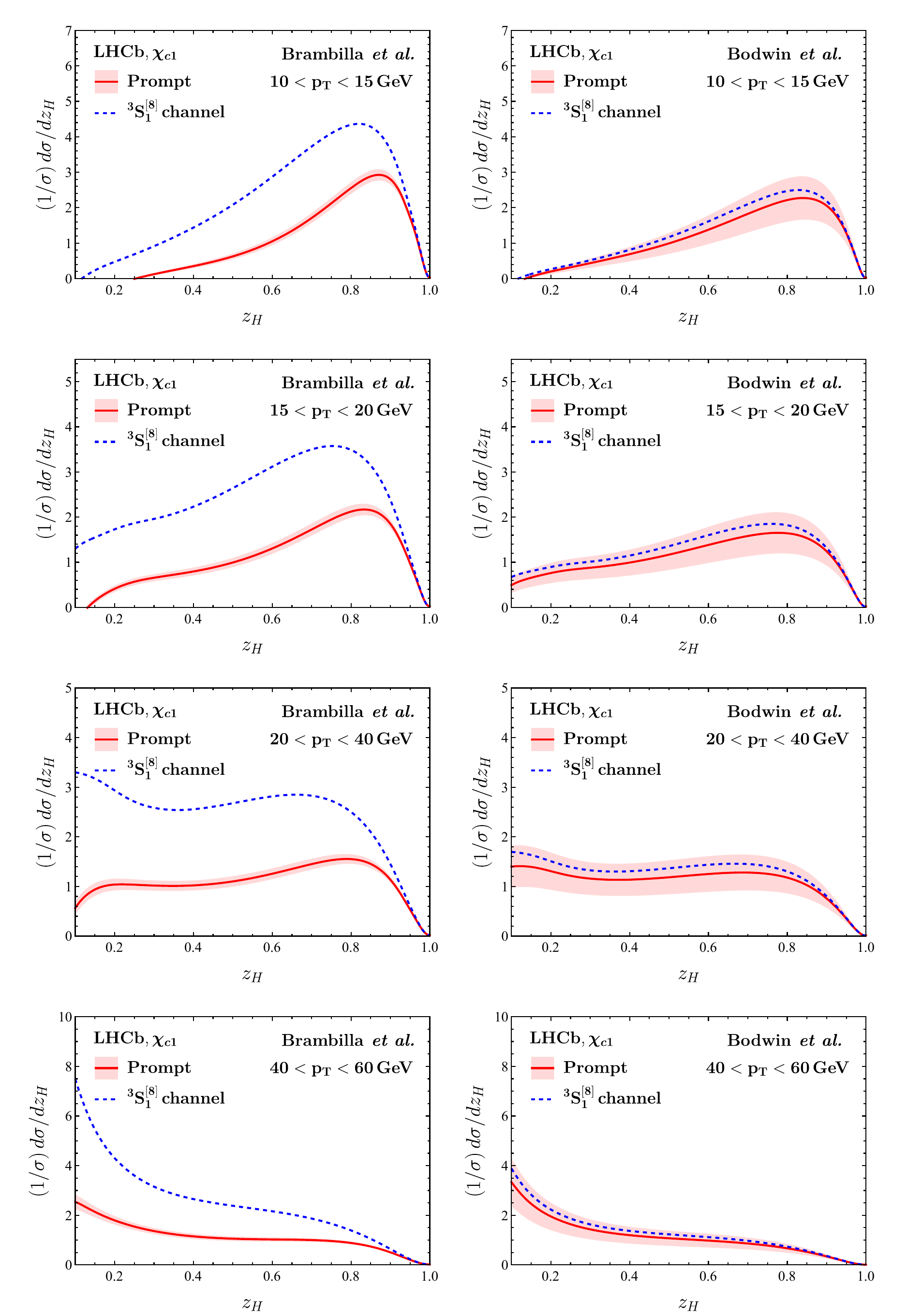}
    \caption{
        Normalized $z_H$ distribution for $\chi_{c1}$ production with the LHCb kinematic selection.
        The left and right columns correspond to the Brambilla and Bodwin LDME sets, respectively, and the rows correspond to increasing $p_T$ bins.
    }
    \label{fig:chic1_LHCb}
\end{figure}

\begin{figure}[h]
    \centering
    \includegraphics[width=0.73\textwidth]{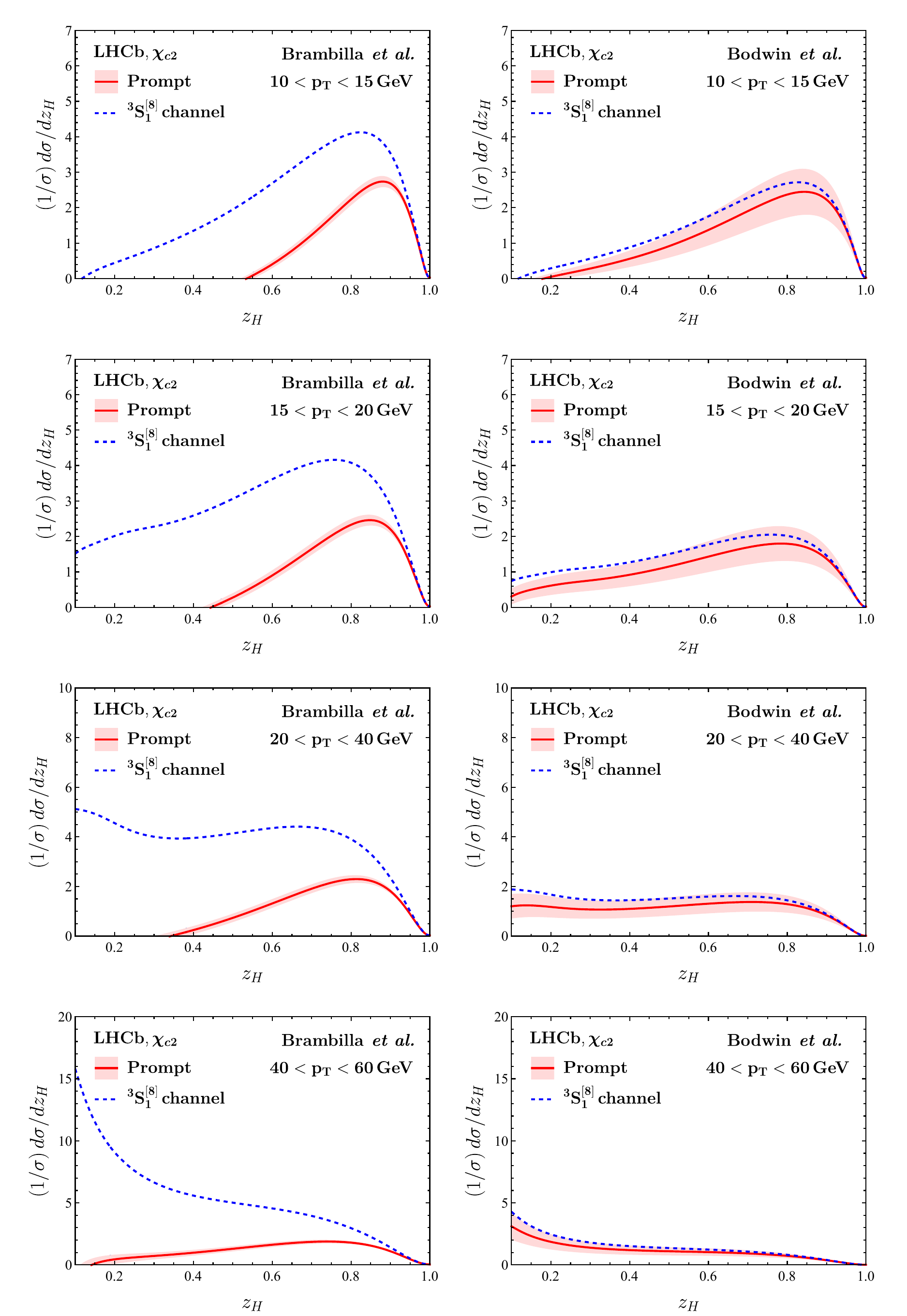}
    \caption{
        Normalized $z_H$ distribution for $\chi_{c2}$ production with the LHCb kinematic selection.
        The left and right columns correspond to the Brambilla and Bodwin LDME sets, respectively, and the rows correspond to increasing $p_T$ bins.
    }
    \label{fig:chic2_LHCb}
\end{figure}

\begin{figure}[h]
    \centering
    \includegraphics[width=0.73\textwidth]{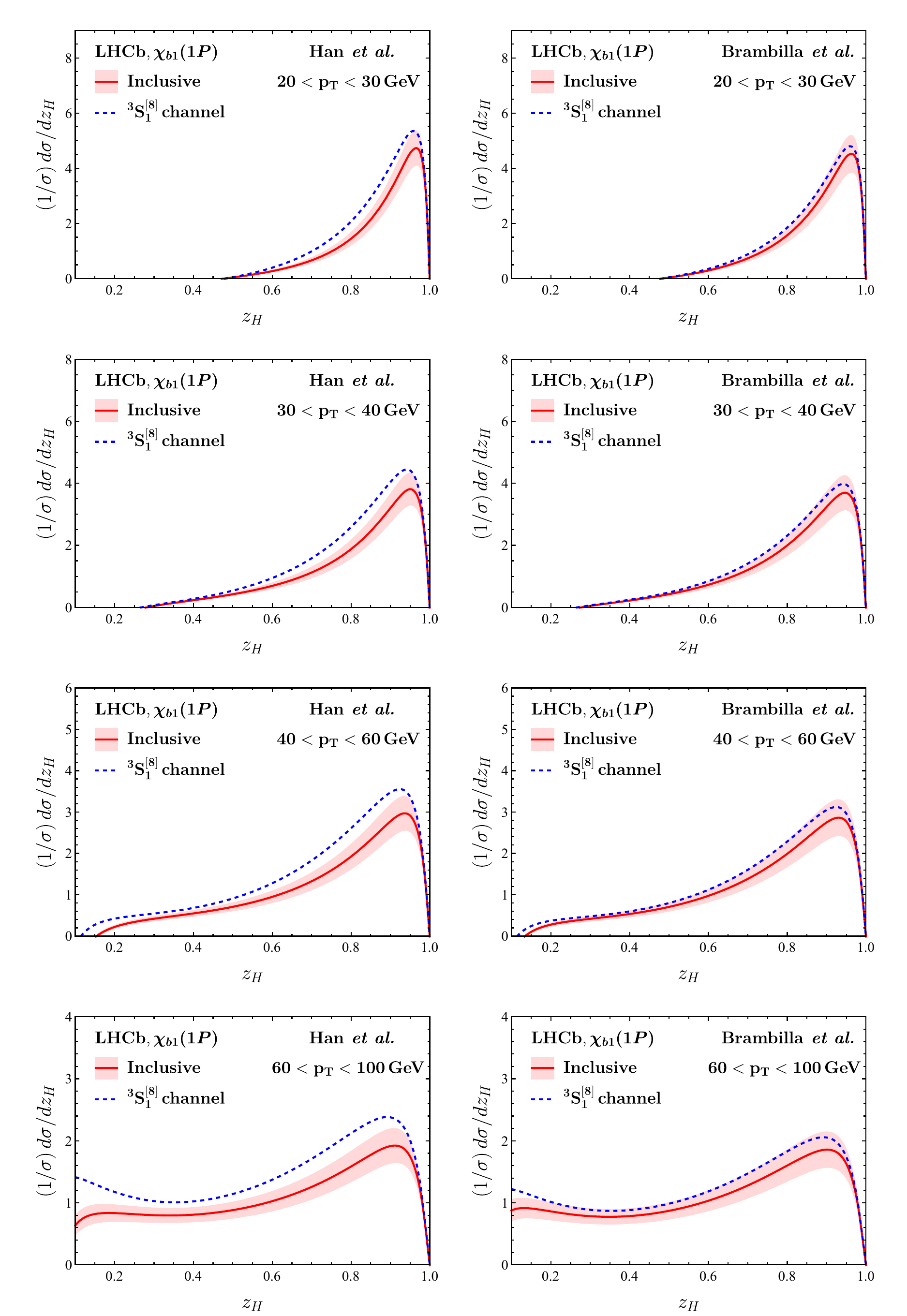}
    \caption{
        Normalized $z_H$ distribution for $\chi_{b1}(1P)$ production with the LHCb kinematic selection.
        The left and right columns correspond to the Han and Brambilla LDME sets, respectively, while the four rows correspond, from top to bottom, to the $p_T$ bins $20$--$30$, $30$--$40$, $40$--$60$, and $60$--$100~\text{GeV}$.
    }
    \label{fig:chib11P_LHCb}
\end{figure}

\begin{figure}[h]
    \centering
    \includegraphics[width=0.73\textwidth]{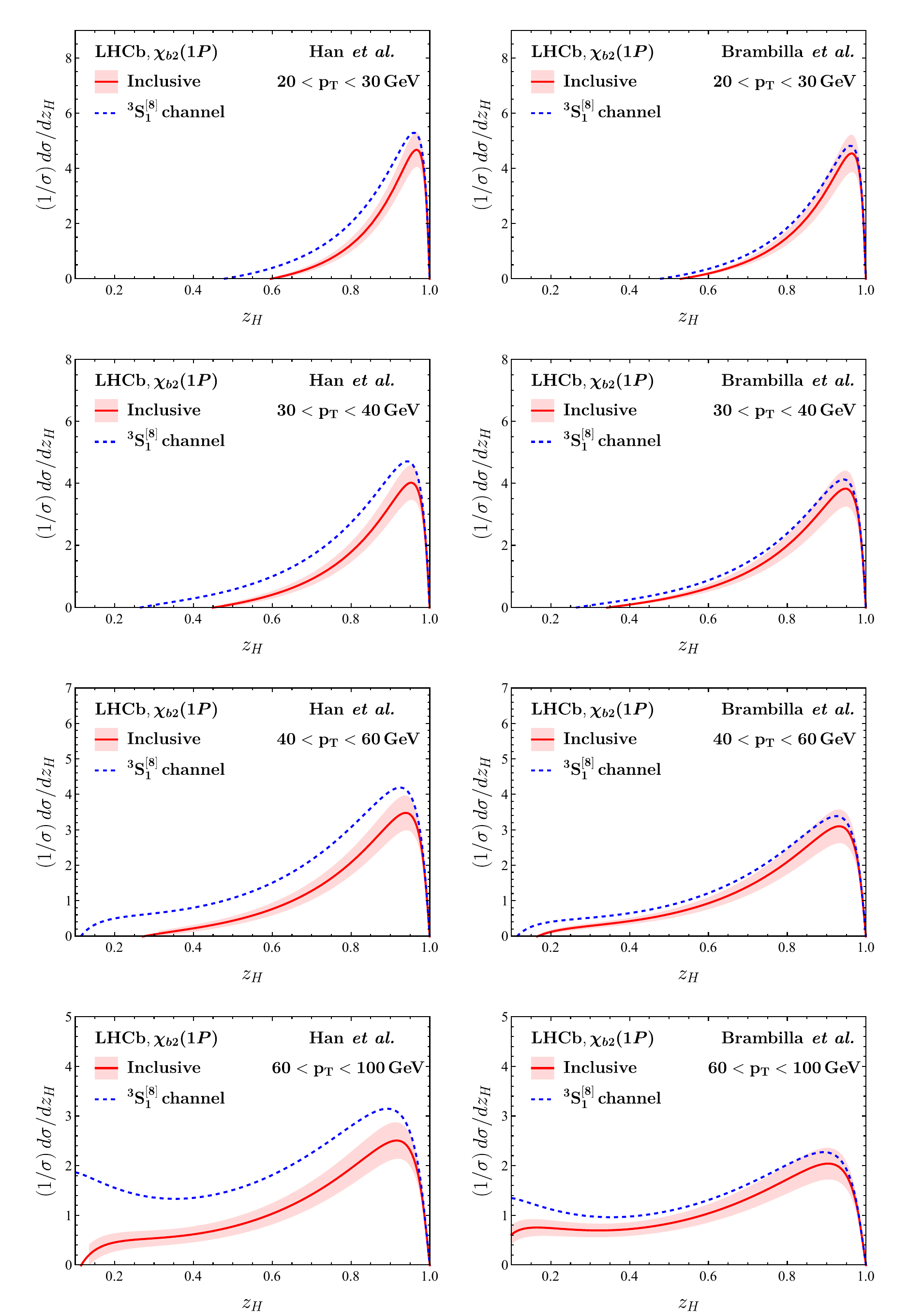}
    \caption{
        Normalized $z_H$ distribution for $\chi_{b2}(1P)$ production with the LHCb kinematic selection.
        The left and right columns correspond to the Han and Brambilla LDME sets, respectively, and the rows correspond to increasing $p_T$ bins.
    }
    \label{fig:chib21P_LHCb}
\end{figure}

\begin{figure}[h]
    \centering
    \includegraphics[width=0.73\textwidth]{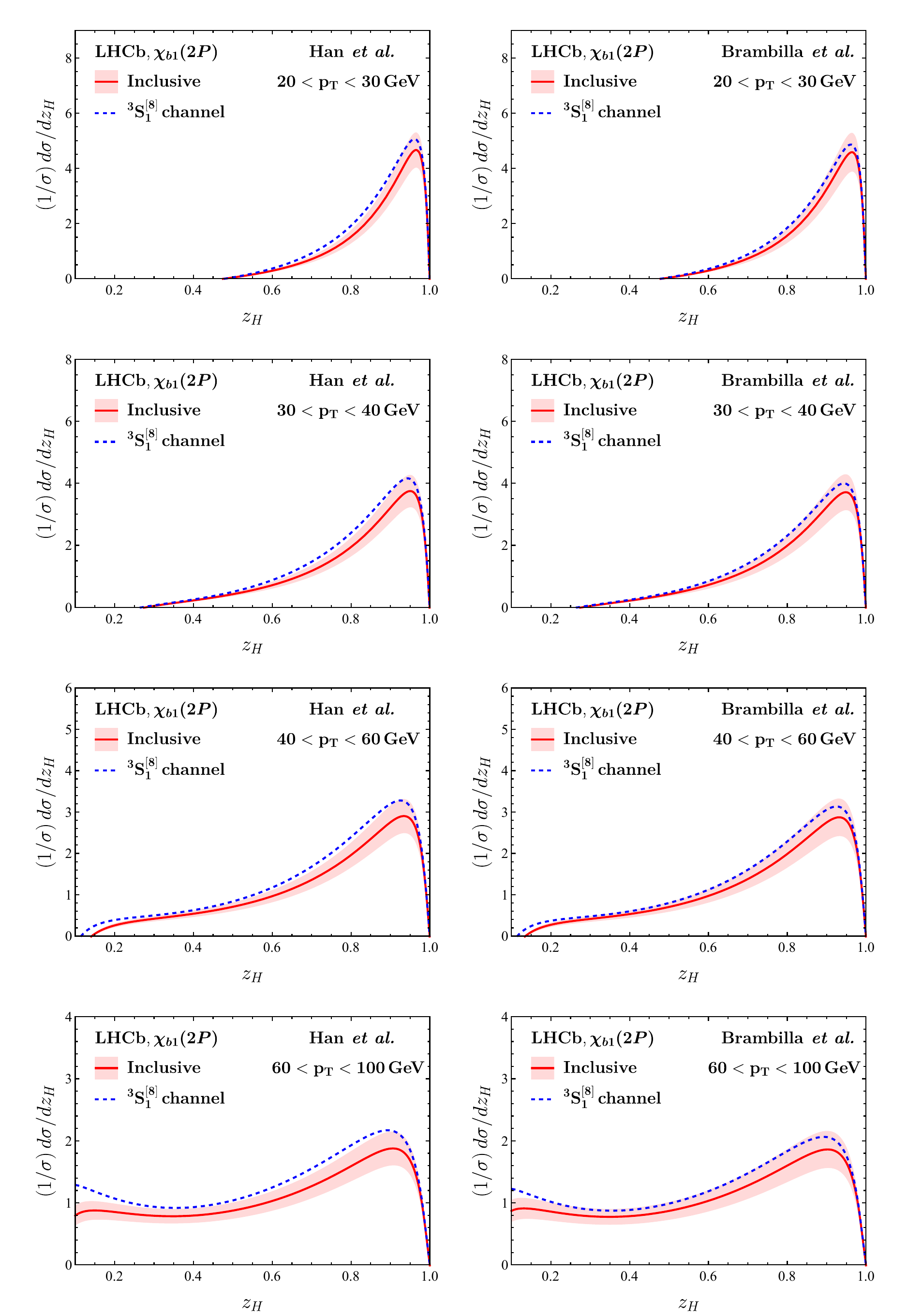}
    \caption{
        Normalized $z_H$ distribution for $\chi_{b1}(2P)$ production with the LHCb kinematic selection.
        The left and right columns correspond to the Han and Brambilla LDME sets, respectively, and the rows correspond to increasing $p_T$ bins.
    }
    \label{fig:chib12P_LHCb}
\end{figure}

\begin{figure}[h]
    \centering
    \includegraphics[width=0.73\textwidth]{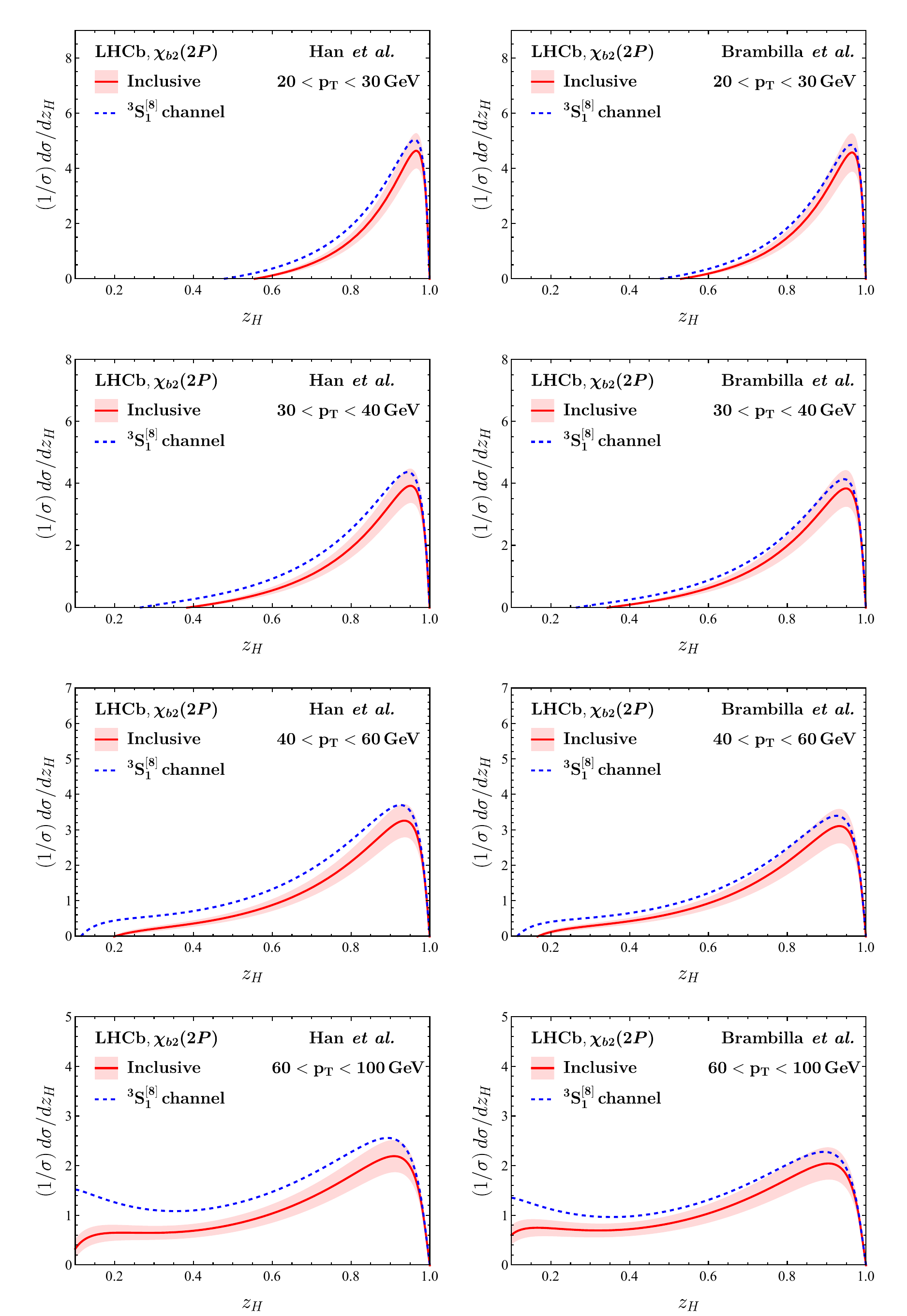}
    \caption{
        Normalized $z_H$ distribution for $\chi_{b2}(2P)$ production with the LHCb kinematic selection.
        The left and right columns correspond to the Han and Brambilla LDME sets, respectively, and the rows correspond to increasing $p_T$ bins.
    }
    \label{fig:chib22P_LHCb}
\end{figure}

\FloatBarrier


\bibliographystyle{JHEP}
\bibliography{FJFbib.bib}

\end{document}